\documentclass[11pt,letterpaper]{article}

\usepackage{amsmath,amssymb,graphicx}
\usepackage{natbib}
\usepackage{color,soul}
\usepackage[margin=2.5cm]{geometry}
\usepackage{topsection}
\usepackage{booktabs}

\newcommand{\pcite}[1]{\citeauthor{#1}'s \citeyearpar{#1}}

\newcommand{\zbf}{{\mathbf{z}}}
\newcommand{\ybf}{{\mathbf{y}}}
\newcommand{\xbf}{{\mathbf{x}}}
\newcommand{\wbf}{{\mathbf{w}}}
\newcommand{\rbf}{{\mathbf{r}}}
\newcommand{\mubf}{{\boldsymbol{\mu}}}
\newcommand{\Sbb}{{\mathbb{S}}}
\newcommand{\half}{{\frac 12}}
\newcommand{\T}{{\scriptscriptstyle\mathsf{T}}}
\DeclareMathOperator*{\argmax}{argmax}
\newcommand{\Mcal}{{\mathcal{M}}}
\DeclareMathOperator{\E}{E}
\DeclareMathOperator{\Var}{Var}

\DeclareMathOperator{\Cov}{Cov}
\newcommand{\ud}{{\,\mathrm{d}}}

\begin{document}
\title{Estimation and prediction for spatial generalized linear mixed models with parametric links via reparameterized importance sampling}
\author{Evangelos Evangelou\,$^{1}$
  and Vivekananda Roy\,$^{2}$\\[5pt]
$^1$ \small{Department of Mathematical Sciences, University of Bath, Bath,
  BA2 7AY, UK.}\\
$^2$ \small{Department of Statistics,
Iowa State University,
3415 Snedecor Hall,
Ames, IA, USA.}
}
\date{\today{}\\
\footnotetext{\textit{Address for correspondence}: Evangelos Evangelou,
  Department of Mathematical Sciences, University of Bath, Bath, BA2 7AY,
  UK. \texttt{email:~ee224{@}bath.ac.uk}}
}

\maketitle

\begin{abstract}
  Spatial generalized linear mixed models (SGLMMs) are popular for
  analyzing non-Gaussian spatial data. These models assume a prescribed
  link function that relates the underlying spatial field with the mean
  response. There are circumstances, such as when the
  data contain outlying observations, where the use of a prescribed link
  function can result in poor fit, which can be improved by using a
  parametric link function. Some popular link
  functions, such as the Box-Cox, are unsuitable
  because they are inconsistent with the Gaussian assumption of the spatial
  field. We present sensible choices of parametric link functions which
  possess desirable properties. It is important to estimate the
  parameters of the link function, rather than assume a known value. To
  that end, we present a generalized importance sampling (GIS) estimator
  based on multiple Markov chains for empirical Bayes analysis of SGLMMs.
  The GIS estimator, although more efficient than the simple
  importance sampling, can be highly variable when used to estimate the
  parameters of certain link functions. Using suitable reparameterizations
  of the Monte Carlo samples, we propose modified GIS estimators that do
  not suffer from high variability. We use
  Laplace approximation for choosing the multiple importance densities in
  the GIS estimator. Finally, we develop a methodology for selecting models
  with appropriate link function family, which extends to choosing a
  spatial correlation function as well. We present an ensemble prediction
of the mean response by appropriately weighting the estimates from different models. The proposed
  methodology is illustrated using simulated and real data examples.

  \mbox{}\\[5pt]
\end{abstract}

\section{Introduction}
\label{sec:intr}
Spatial generalized linear mixed models (SGLMMs), introduced by
\cite{digg:tawn:moye:1998}, are often used for analyzing non-Gaussian
spatial data that are observed in a continuous region \cite[see
e.g.][]{zhan:2002, chri:waag:2002, digg:ribe:chri:2003, chri:2004}. SGLMMs
are generalized linear mixed models where the random effects consist of a spatial
process. Conditional on the spatial process, the response variables are
assumed to follow a distribution which only depends on the site-specific
conditional means. A link function relates the means of the response
variable to the underlying spatial process. For the binomial response variable, a logit
or probit link is often assumed, while for the Poisson distribution, a logarithmic
link is used. It has been recently shown that the use of a flexible parametric family of link functions (instead of a
known fixed link) may produce better inference and prediction
\citep{chri:2004, roy:evan:zhu:2016}.

Parametric links have been discussed in the literature of generalized
linear models (GLMs). For the binomial GLM, for modeling dose-response curves,
\cite{prentice1976generalization} introduces a two-parameter link
function given by the
quantile of the logarithm of an $F$-distributed random variable, also
called the type~IV generalized logistic distribution
\citep{johnson1995continuous2}. This link function includes the logit
and probit links as special cases.  \cite{liu:2004},
\cite{koenker2009parametric}, and \cite{roy:2014} discuss the link
function defined by the quantile of the Student's $t$ distribution,
the so-called robit link, which approximates the logit and probit
links but provides robust inference in the presence of outlying
observations. \cite{wang2010generalized} use the extreme-value
quantile link function which is non-symmetric and can therefore be
used when the rate of change in the success probability approaches 0
at a different rate than it approaches 1. Other authors discussing
parametric links for binary data include \cite{Aranda81, GuerrJohn82,
  Stukel, nagler1994scobit, chen1999new} and \cite{bazan2006}. For
Poisson data, \cite{basu2005estimating} use a Box-Cox link
function.

The added flexibility of parametric links introduces the complexity of
having to estimate the parameters of the link function. 
In general, for SGLMMs, the likelihood function can be written only as a
multi-dimensional integral and does not have a closed form expression. One
way to approximate the intractable likelihood in SGLMMs is by importance
sampling \citep{chri:2004}. Samples are generated from an importance
sampling distribution which are then used for approximating the likelihood
by calculating Monte Carlo (MC) averages. The accuracy of the approximation
depends on the choice of the importance sampling distribution which can be
difficult to elicit if one has to estimate the likelihood for a wide range
of parameter values. Generalized importance sampling (GIS) is an efficient
importance sampling methodology based on multiple proposal (importance)
densities for estimating the ratios of marginal likelihoods for SGLMMs.
These ratios of marginal likelihoods are called Bayes factors (BFs). If the
marginal likelihood in the denominator (of BFs) is fixed at a parameter value,
while the parameter in the numerator is allowed to vary, then maximization
of the BFs is equivalent to maximization of the marginal likelihoods
resulting in the empirical Bayes (EB) estimate.
\cite{roy:evan:zhu:2015,roy:evan:zhu:2016} used this idea to estimate not
only the link parameter but other parameters as well, such as the spatial
range and relative nugget. One benefit of using the EB methodology over a
fully Bayesian approach is that it avoids having to specify a prior for
these parameters as prior elicitation for these parameters is often
difficult, and improper priors on these parameters generally lead to
improper posteriors \citep{berg:deol:sans:2001, chri:waag:2002}. Also in
case of a fully Bayesian analysis, the Markov chain Monte Carlo (MCMC)
algorithms may suffer from slow mixing \citep{chri:2004, roy:2014}.

In this paper we use an EB methodology, implemented by an efficient
GIS based on reparameterizations of the MC samples, to fit SGLMMs with
parametric links.  The contributions of the paper are in four areas:
\begin{itemize}
\item \textbf{Link functions suitable for spatial data analysis.}
Despite the abundance of parametric link functions in the literature, not every
link function is suitable for spatial data analysis, where the link function
relates the mean response to the latent spatial field. Because the latent
spatial field is assumed to be a Gaussian process, it is required that the
link function maps \textit{onto} the whole real line. Otherwise, this
creates an inconsistency in the model because not every possible value of
the Gaussian process can correspond to a mean value in the distribution of
the observations.

Some popular link functions discussed in the literature,
including the Box-Cox link, do not satisfy this requirement. This fact was
noted by \cite{chri:2004} in the case of the Poisson Box-Cox model. 
In this paper we provide  modifications of these links, by smoothing
transitions to their limits, which inherit
their flexibility, but also are consistent with the SGLMM. These link
functions have not been proposed before in the literature, even for traditional
GLMs.

\item \textbf{Improved GIS estimators via reparameterization and control variates.}
When approximating integrals numerically, a suitable change-of-variables
can improve numerical stability. For importance sampling integration, this
corresponds to transforming the MC samples. It has been shown that
reparameterizations can drastically improve mixing of Gibbs samplers
\cite[see e.g.][]{simp:niem:roy:2016, roy:2014, vand:meng:2001,
  liu:wu:1999}. We show in this paper how the GIS estimator {\itshape
  without} transformation of \cite{roy:evan:zhu:2016} can produce biased
estimates. We then discuss how to choose
suitable transformations to produce better estimators. Thus we derive
modified GIS estimators based on transformed (reparameterized) samples.
Because of the additional computational cost of transforming the MC
samples, some transformations can be slow. In this case, we show how a
different, suitable transformation can produce accurate results in less computational
time. We also use the proposed transformations to improve the performance
of \pcite{geye:1994} reverse logistic regression estimator. Although \cite{chri:2004}
suggested the use of the mean transformation for the \textit{simple}
importance sampling estimator, this paper is the first to present
\textit{generalized} importance sampling estimators based on
\textit{general} transformations.

Another approach for reducing the variability of IS estimates is the use of
control variates \citep{owen:zhou:2000}. \cite{doss:2010} used control
variates to reduce the variability of BF estimates for multiple IS
estimators. We show in this paper how the approach of \cite{doss:2010} can
be applied to the reparameterized GIS estimators we propose.

\item \textbf{Selection of proposal distributions using Laplace
  approximation.}
The performance of any IS estimator, including GIS, crucially depends on
the proposal (importance) distributions. In the literature, there is no
systematic method available for selecting these proposal distributions,
although it has been generally deemed as difficult
\citep{buta2011computational}. Use of good importance densities is
particularly important for spatial models due to potential multimodality of
the likelihoods \citep{mard:1989}. Choosing representative importance
sampling distributions can be very difficult if there are too many
parameters to estimate. In this paper we use Laplace approximation to
integrate out the latent spatial field and thus derive an approximation to
the marginal likelihood of the observed data. This approximation is used to
choose ``good'', representative importance sampling distributions.

\item \textbf{Model selection.}
A typical problem faced by practitioners is the choice among
different spatial correlation families and the choice of the link function.
An established measure of model comparison and weighting is AIC. However,
calculation of AIC is not straightforward for SGLMMs as the likelihood is
intractable. This paper is the
first to address the problem of spatial model selection using GIS.
We demonstrate how the GIS estimator developed in this paper
can be used to approximate the AIC by evaluating the Bayes factors between
the candidate models. The approximated AIC can be used for model selection and
weighting in the spirit of \cite{buckland1997model}, thus providing
ensemble estimation and prediction methods.
\end{itemize}

The remainder of the paper is organized as follows. In
Section~\ref{sec:sglm} we discuss the SGLMM, and present some suitable link
functions for binomial and Poisson/gamma models. In Section~\ref{sec:eb} we
develop the estimation methodology and a method for selection of
importance densities. This section also contains a measure of comparison
between models with different link and correlation function families. In
Section~\ref{sec:exam} we use simulation studies to demonstrate the issues
with importance sampling and how our modified methods based on
transformation can address these. We also demonstrate the performance of
the proposed model selection criterion via a different simulation study.
The methods discussed in this paper are applied to two real-data examples
in Section~\ref{sec:examples}. Finally, Section~\ref{sec:disc} presents the
conclusions of this paper. Some technical derivations are relegated to
Appendix~\ref{sec:appendixA}. A summary of the steps involved in the
proposed computational and inferential procedure is presented in
Appendix~\ref{sec:summ}. Finally, Appendix~\ref{sec:list-skeleton-points}
contains further details about the examples.

\section{Spatial generalized linear mixed models}
\label{sec:sglm}

Let $\{Z(s), s \in \Sbb\}$ be a Gaussian random field with mean
function $E(Z(s)) = \sum_{j=1}^p x_j(s) \beta_j$, where
$\beta=(\beta_1, \dots, \beta_p)' \in \mathcal{R}^p$ are the unknown
regression parameters, $\xbf(s) = (x_1(s),\dots, x_p(s))$ are the
known location dependent covariates, and the covariance function
$\Cov(Z(s), Z(s')) = \sigma^2 \rho_\theta(s, s') + \tau^2
I_{\{s=s'\}}$.  Here $\rho_\theta(s, s')$ is the spatial correlation
function which models the dependence between distinct locations. In
this paper we assume a stationary and isotropic correlation, i.e.
$\rho_\theta(s, s') = \rho_\theta(\|s-s'\|)$, where $\| s-s' \|$
denotes the Euclidean distance between $s$ and $s'$. Some examples of
correlation functions are the \textit{Mat\'{e}rn}, the
\textit{exponential-power}, and the \textit{spherical} parametric
families \citep{digg:ribe:chri:2003}. These functions depend on
parameters $\theta$. In the case of the spherical family, there is
only one parameter, the spatial range $\phi$, i.e. $\theta = \{\phi\}$,
but in the case of the Mat\'ern and exponential-power families, there
is respectively an additional smoothness or power parameter $\kappa$,
i.e.  $\theta = \{\phi,\kappa\}$. The parameter $\sigma^2$ is called
the partial sill, and $\tau^2$ is called the nugget effect. The nugget
effect can be interpreted as micro-scale variation, measurement error,
or a combination of both. It is convenient to let
$\omega = \tau^2/\sigma^2$ and write the covariance as
$\Cov(Z(s), Z(s')) = \sigma^2 [\rho_\theta(s, s') + \omega
I_{\{s=s'\}}]$.

Conditional on the realized value of the Gaussian random field,
$\{z(s), s \in \Sbb\}$, the response/observation process
$\{Y(s), s \in \Sbb\}$ is assumed to consist of independent random
variables, and for each $s \in \Sbb$ the distribution of
$Y(s) | z(s)$ has conditional mean
\[ \E(Y(s) | z(s)) = t(s) \mu(s), \]
where $t(s)$ is a known function and $\mu(s)$ is related to $z(s)$
through a link function $h_\nu$ such that
\begin{equation}
  \label{eq:1}
  h_\nu(\mu(s)) = z(s).
\end{equation}
The Gaussian random field is unobserved while the response process is
observed at a finite set of locations $s_1,\ldots,s_n \in \Sbb$.
We write $y_i = y(s_i)$, $\mu_i = \mu(s_i)$ and so on.

The link function $h_\nu$ is assumed to belong to a parametric family
depending on parameters $\nu$. The conditional model for the observation
process depends on $z(s)$ only through its relationship with $\mu(s)$ and
can be written as
\begin{equation*}
  p[y(s)|z(s);\nu] = p[y(s)|\mu(s) = f_\nu(z(s))] ,
\end{equation*}
where we use $p[\cdot]$ to denote the pmf/pdf of the enclosed expression.
We also use $f_\nu (\cdot) = h^{-1}_\nu (\cdot)$ to denote the inverse of the link function.

We now present two examples of SGLMMs appropriate for binary and count
data respectively.
\cite{roy:evan:zhu:2016} consider the following robust SGLMM for analyzing
spatial binomial data. For any $s_1,\ldots,s_n \in \Sbb$,
conditional on $\{z(s)\}$, the response variables $Y_1,\ldots,Y_n$ are
assumed to follow
$Y_i | z_i \stackrel{\mathrm{ind}}{\sim} \mathrm{Binomial}(t_i,
\mu_i)$ with $\mu_i = G_{\nu} (z_i)$,
where $G_\nu (\cdot)$ is the cumulative
distribution function (cdf) of the standard Student's $t$ distribution with
degrees of freedom $\nu$ and $t_i$ is a known constant (number of trials
at the location $s_i$) for $i=1, \dots, n$. This model is called the
spatial robit model because it is more robust to outlying observations
compared to the standard logistic and probit models. 

Our second example is used to analyze spatial count data, where
$Y_i | z_i
\stackrel{\mathrm{ind}}{\sim}  \mathrm{Poisson} (t_i\mu_i)$, with 
$\mu_i =
h_\nu^{-1} (z_i)$. Here $t_i$ may represent the
length of the recording period over which $y_i$ is observed,
or the area within which $y_i$ is counted. \cite{chri:2004} considers the
Box-Cox family of link functions given by
\begin{equation}
  \label{eq:2}
  h_\nu(\mu_i) = \begin{cases}
    (\mu_i^\nu - 1)/\nu, & \text{if $\nu \neq 0$,} \\
    \log (\mu_i), & \text{if $\nu =0$.}
  \end{cases}
\end{equation}
So the commonly used log link function, $h(\mu_i)= \log (\mu_i)$ is a
special case of the above Box-Cox family of link functions.  For
analyzing a data set of radionuclide concentrations on Rongelap
island, \cite{chri:2004} provides evidence that the log-link, as used
by \cite{digg:tawn:moye:1998}, may not be a good choice and  uses
the above Box-Cox family of link functions.

A problem with the Box-Cox link is that it is inconsistent with the SGLM
model for $\nu \neq 0$ because it imposes the restrictions $z_i > -1/\nu$
and $z_i < -1/\nu$ if $\nu > 0$ or $\nu < 0$ respectively, which contradicts
the Gaussian assumption for $z_i$. To avoid this issue, \cite{chri:2004}
extended the model to allow for $z_i \in \mathcal{R}$ such that
$z_i \in (-\infty, -1/\nu] \Leftrightarrow \mu_i = 0$ if $\nu > 0$, and
$z_i \in [-1/\nu, \infty) \Leftrightarrow \mu_i = 0$ if $\nu < 0$, and
$\mu_i = 0 \Rightarrow y_i = 0$ with probability 1. However in this case
the link function is not invertible.

\subsection{Parametric link functions}
\label{sec:para}

We now discuss some desirable properties of link functions.
To facilitate inference, we require the function to be
monotone and differentiable. In order to be consistent with the SGLMM,
we require that the function maps the range of values for the mean (of
the observation process) \textit{onto} the real line. This property is not
satisfied, for example by the Box-Cox link used in \cite{chri:2004}
when $\nu \neq 0$. We present below some suggestions for parametric
links for different models.

\subsubsection{Binomial response variables}
\label{sec:bin}
For binomial response variables, the mean, $f_\nu (z)$, lies between 0 and 1.
It is helpful to think of the inverse link function as having the form
$f_\nu (z) = F_\nu(z)$ where $F_\nu$ is the cdf of a real-valued continuous random variable
with support being the whole real line. The popular logistic
and probit models are derived by letting $F_\nu$ be the cdf of the logistic
and standard normal distributions respectively, while the robit link of
\cite{liu:2004} corresponds to the cdf of the standard Student's $t_\nu$
distribution.

\cite{roy:evan:zhu:2016} demonstrate the advantages of using a
parametric link function for robust spatial inference under model
misspecification, or in the presence of outlying observations. In the
latter case, the robit link function with low degrees of freedom would
be more appropriate choice.  Similar behavior can be achieved by using
a computationally efficient approximation to the $t_\nu$ cdf due to
\cite{wallace}, that is,
\begin{equation}
  \label{eq:6}
  F_\nu(z) = \Phi(\zeta),\ \zeta = \mathrm{sign}(z)
                     \frac{8\nu+1}{8\nu+3} \sqrt{\nu\log(1+z^2/\nu)},
\end{equation}
where $\Phi(\cdot)$ denotes the cdf of the standard normal distribution.

In other situations, one may want to use a non-symmetric cdf, if e.g. the
rates at which the success probability 
approaches 0 and 1 are different. The
generalized extreme value (GEV) link was proposed by \cite{WangDeyBan:10}
and can be used for this purpose. This link corresponds to
\begin{equation}
  \label{eq:3}
  F_\nu(z) =
  \begin{cases}
    \exp \left\{-\max(0, 1 + \nu z)^{-1/\nu}\right\}, & \text{if $\nu \neq
      0$,} \\
    \exp \left\{-\exp (-z)\right\}, & \text{if $\nu = 0$,}
  \end{cases}
\end{equation}
which puts restrictions on $z$ as $z > -1/\nu$ if $\nu > 0$ and
$z < -1/\nu$ if $\nu <0$. A link that behaves similarly as
\eqref{eq:3} but maps onto the real line can be obtained by
letting $f_\nu (z) = F_\nu (z)$ where
\begin{equation}
  \label{eq:4}
  F_\nu(z) =
  \begin{cases}
    \exp \left\{-(1 + |\nu| |z|)^{-\mathrm{sign}(z)/|\nu|} \right\}, & \text{if
      $\nu \neq 0$,} \\
    \exp \left\{-\exp (-z)\right\}, & \text{if $\nu = 0$.}
  \end{cases}
\end{equation}
We will refer to the link corresponding to~\eqref{eq:4} as the
\emph{modified GEV link}. Since this link function depends only on $|\nu|$,
either $\nu \in (-\infty,0]$, or $\nu \in [0,\infty)$ is assumed.

One advantage of the standard GEV link is that it allows for positive as
well as negative skewness while the modified GEV link only allows for
positive skewness. This means that the modified GEV only considers the case
where the probability of success approaches 1 faster than it approaches 0,
and may result in poor fit for some data. However, for every cdf
$F_\nu(z)$, $F_\nu^*(z) = 1-F_\nu(-z)$ is a also a cdf, and if $F_\nu(z)$
generates a
positively skewed link, then $F^*_\nu(z)$ generates a negatively
skewed link. This is
equivalent to interpreting successes as failures and vice versa. When
$F_\nu(z)$ is the cdf in~\eqref{eq:4} then $F^*_\nu(z)$ corresponds to the
\emph{negatively-skewed} modified GEV link, which includes the popular
complementary log-log link as a special case.

\subsubsection{Poisson and gamma response variables}
\label{sec:pois}
For Poisson and gamma models the mean response can take any positive real value.
 A general family of inverse link
functions can be derived by
\begin{equation*}
  f_\nu(z) = -\log F_\nu(-z), 
\end{equation*}
where $F_\nu(z)$ is a cdf as in the binomial case. For instance the choice
$f_\nu(z) = -\log F_{-\nu}(-z)$ where $F_\nu$ is the GEV cdf
in~\eqref{eq:3} produces the Box-Cox link~\eqref{eq:2} with the
logarithmic link as a special case. Thus a modified Box-Cox link can be
derived by using the modified GEV cdf (given in \eqref{eq:4}) as
\begin{equation*}
  h_\nu(\mu) =
  \begin{cases}
    \frac{\mu^\nu-1}{\nu} & \text{if $\nu > 0$ and $\mu \geq 1$}, \\
    \frac{1-\mu^{-\nu}}{\nu} & \text{if $\nu > 0$ and $\mu < 1$}, \\
    \log \mu & \text{if $\nu=0$} .
  \end{cases}
\end{equation*}

\section{Empirical Bayes estimation of SGLMMs}
\label{sec:eb}

Suppose that the data $\ybf = (y_1, \dots, y_n)$ consist of a single
realization of the process $\{Y(s), s \in \Sbb\}$ mentioned in
Section~\ref{sec:sglm} at known sampling locations
$s_1, \dots, s_n \in \Sbb$. Let us divide all unknown parameters into
two categories $\psi \equiv (\beta, \sigma^2)$ and
$\xi \equiv (\nu, \theta)$ depending on whether a conjugate prior for
those parameters given $\zbf$ exists or not respectively. One of the reasons for
this split is that it is straightforward to sample from the full
conditionals of the parameters in $\psi$, as these are standard distributions, but not so if we had
assumed a prior for $\xi$.
The
likelihood function of SGLMM is not available in closed form, but only
as a high dimensional integral, that is,
\begin{equation}
  \label{eq:likgls}
  L_\xi(\psi | \ybf) \equiv L(\psi, \xi | \ybf)
  = \int_{\mathcal{R}^n} p[\ybf, \zbf | \psi, \xi] \ud \zbf
  =\int_{\mathcal{R}^n} p[\ybf|\zbf,\nu] p[\zbf | \psi, \xi] \ud \zbf,
\end{equation}
where $\zbf = (z_1, \dots, z_n)$, $z_i \equiv z(s_i)$,
$p[\ybf|\zbf,\nu] = \prod_{i=1}^n p[y_i |z_i, \nu]$ with
$p[y_i |z_i, \nu] = p[y_i|\mu_i = f_\nu(z_i)]$ being the conditional
density of $y_i| z_i$, and $p[\zbf | \psi, \xi]$ is the multivariate
Gaussian density for $\zbf$ with mean vector $X\beta$ and covariance matrix
involving the parameters $\sigma^2$ and $\theta$, and $X$ is the known
$n\times p$ matrix defined by $X_{ij} = x_j(s_i)$.

Note that the Gaussian prior for $\beta$ (conditional on $\sigma^2$) and
scaled inverse chi-square prior for $\sigma^2$ are conjugate priors for
$\psi =(\beta, \sigma^2)$ for the joint density
$p[\ybf, \zbf | \psi, \xi]$. Let $\pi(\psi)$ be the prior on $\psi$
obtained from assuming $\beta | \sigma^2 \sim N(m_b, \sigma^2\, V_b)$, and
$\sigma^2 \sim \chi^2_{ScI} (n_\sigma, a_\sigma)$ where the hyperparameters
$m_b, V_b, a_\sigma, n_\sigma$ are assumed known.

Consider the augmented joint density $p[\ybf, \zbf | \psi, \xi]$ and the
corresponding so-called {\it complete} posterior density
\begin{equation}
  \label{eq:compost}
  \pi_{\xi}( \psi, \zbf| \ybf) = \frac{p[\ybf, \zbf | \psi, \xi]
    \pi(\psi)}{m_{\xi}(\ybf)},
\end{equation}
where
\begin{equation}
  \label{eq:marg}
  m_{\xi}(\ybf) = \int_{\mathcal{R}^p \times \mathcal{R}_+}
  \int_{\mathcal{R}^n} p[\ybf, \zbf | \psi, \xi] \pi(\psi)
  \ud \zbf \ud\psi =
  \int_{\mathcal{R}^p \times \mathcal{R}_+} L_\xi(\psi | \ybf) \pi(\psi)
  \ud\psi
\end{equation}
is the normalizing constant (also known as the marginal density). The
empirical Bayes (EB) approach to inference is to estimate $\xi$ by
maximizing this marginal density. Suppose $\hat{\xi}$ is the
maximizer, i.e, $\hat{\xi} = \argmax m_\xi (\ybf)$. Then, the
posterior density $\pi_{\hat\xi}( \psi, \zbf| \ybf)$ of $(\psi,\zbf)$,
conditioned on the observed data $\ybf$ and $\xi = \hat{\xi}$ is used
to infer about $(\psi,\zbf)$. Typically, for fixed $\xi$, one would
sample iteratively from the full conditionals
$\pi_{\xi}(\psi|\zbf, \ybf)$ and $\pi_{\xi}(\zbf|\psi, \ybf)$ to run a
Gibbs sampler. Since we use conjugate priors for $\psi$, sampling from
the former is straightforward, while for the latter a
Metropolis-Hastings algorithm is used as in
\cite{digg:tawn:moye:1998}.

Note that for
any arbitrary fixed $\xi_1$, $\hat{\xi}$ is equal to $\hat{\xi}$ =
argmax $B_{\xi, \xi_1}$ where
$B_{\xi, \xi_1} = m_{\xi}(\ybf)/ m_{\xi_1}(\ybf)$ is the BF
for the model indexed by $\xi$ relative to the model indexed by
$\xi_1$. The reason for considering the latter is that it is often
much easier to compute the ratio $B_{\xi, \xi_1}$ instead of the
marginal likelihood $m_{\xi}(\ybf)$ directly. (Note that in order to
find the maximizer of $B_{\xi, \xi_1}$ we may need to estimate
$B_{\xi, \xi_1}$ for many values of $\xi$.) For example if
$\{\psi^{(i)}, \zbf^{(i)}\}_{i=1}^N$ is a {\it positive Harris} Markov
chain with stationary density $\pi_{\xi_1}( \psi, \zbf| \ybf)$, then a
consistent estimator of $B_{\xi, \xi_1}$ is given by
\begin{equation}
  \label{eq:bf}
  \frac{1}{N} \sum_{i=1}^N \frac{p[\ybf, \zbf^{(i)} | \psi^{(i)}, \xi]}{p[\ybf,
    \zbf^{(i)} | \psi^{(i)}, \xi_1]} \stackrel{\mbox{a.s.}}{\longrightarrow}
  \int_{\mathcal{R}^n} \int_{\mathcal{R}^p \times \mathcal{R}_+} \frac{p[\ybf,
    \zbf | \psi, \xi]}{p[\ybf, \zbf | \psi, \xi_1]} \pi_{\xi_1}(\psi, \zbf| \ybf)
  \ud\psi d\zbf
  =\frac{m_{\xi}(\ybf)}{m_{\xi_1}(\ybf)},
\end{equation}
as $N \rightarrow \infty$, where $p[\ybf, \zbf | \psi, \xi]$ is the
joint density given in \eqref{eq:likgls}. The simple importance sampling (IS)
estimator \eqref{eq:bf} is often unstable as some of the
terms (ratios of densities) take very large values
especially when $\xi$ is not ``close'' to $\xi_1$ \citep{geye:1996,
  chri:2004, doss:2010}.

We now describe the GIS method for estimating $\hat{\xi}$.
A more efficient method for estimating $B_{\xi, \xi_1}$ for a wide
range of values for $\xi$ was proposed initially by \cite{geye:1994}
\cite[see also][]{geye:thomp:1992} and subsequently used by
\cite{doss:2010} and \cite{roy:evan:zhu:2016} among others. The idea
is to choose a \emph{skeleton set} $\Xi = \{\xi_1,\ldots,\xi_k\}$ with
multiple $\xi$ values and generate a Markov chain
$\{\psi^{(j;l)}, \zbf^{(j;l)}\}_{l=1}^{N_j}$ with stationary density
$\pi_{\xi_j}(\psi,\zbf|\ybf)$ for each $j=1,\ldots,k$ and use the
following generalized IS (GIS) estimator
\begin{equation}
  \label{eq:dhatbfsg}
  \hat{B}_{\xi, \xi_1} (\hat {\rbf}) = \sum_{j =1}^k \sum_{l =1}^{N_j}
  \frac{p[\ybf, \zbf^{(j;l)} |\psi^{(j;l)}, \xi]}{\sum_{i =1}^k N_i p[\ybf,
    \zbf^{(j;l)} | \psi^{(j;l)}, \xi_i]/ \hat{r}_i},
\end{equation}
where $\hat {\rbf} = (\hat{r}_1, \hat{r}_2, \dots, \hat{r}_k)$ is the
``reverse logistic regression'' (RL) estimator \citep{geye:1994} of
${\rbf} = (r_1, r_2, \dots, r_k)$ with
$r_i \equiv m_{\xi_i}(\ybf)/m_{\xi_1}(\ybf)$ for $i =2, \dots k$ and
$\hat{r}_1 = 1 = r_1$. This leads to a numerically stable IS estimator with
smaller variance than~\eqref{eq:bf}.

In
order to describe \pcite{geye:1994} RL estimation of ${\rbf}$, define
\begin{equation}
  \label{eq:13}
  \delta_j = -\log r_j + \log \frac{N_j}{N} \;\; \mbox{ for}\; j=1,\dots, k,
\end{equation}
where $N=\sum_{j=1}^k N_j$. The RL estimator of
$\delta = (\delta_1,\dots, \delta_k)$ (and hence of ${\rbf}$) is
obtained by maximizing the log quasi likelihood function
\begin{equation}
  \label{eq:14}
\sum_{j =1}^k \sum_{l =1}^{N_j} \log \tilde{p}_j(\psi^{(j;l)}, \zbf^{(j;l)}; \delta)\;\;\mbox{ with the constraint} \;\;\sum_{j=1}^k \delta_j = 0,
\end{equation}
where
\begin{equation}
  \label{eq:ptil}
  \tilde{p}_j(\psi, \zbf; \delta) = \frac{p[\ybf, \zbf|\psi, \xi_j]e^{\delta_j}}{\sum_{t =1}^k p[\ybf, \zbf | \psi, \xi_t]e^{\delta_t}}.
\end{equation}
Note that, $\tilde{p}_j(\psi, \zbf; \delta)$ is the probability that
$(\psi, \zbf)$ came from the $j$th density
$\pi_{\xi_j}(\psi,\zbf|\ybf)$ given that it belongs to the pooled
sample $\{\psi^{(j;l)}, \zbf^{(j;l)},l=1,\dots,N_j, j=1,\dots,k\}$.
The reason for the constraint in~\eqref{eq:14} is because the
$\delta_i$'s are only identifiable up to a constant, i.e., adding a fixed
constant to~\eqref{eq:13} does not change~\eqref{eq:ptil}. This
unidentifiability is not an issue for us because we only need to estimate
$k-1$ ratios $r_j, j=2,\dots,k$.

\cite{doss:2010} proposed a two stage scheme for using the GIS
estimator \eqref{eq:dhatbfsg}. In the 1st stage based on samples
$\{\psi^{(j;l)}, \zbf^{(j;l)}\}_{l=1}^{\tilde{N}_j}$ from
$\pi_{\xi_j}(\psi,\zbf|\ybf)$, $j=1,\ldots,k$, ${\rbf}$ (the ratios
of marginal likelihoods at $k$ skeleton points) is estimated by the RL method. Then
independent of stage I, new samples
$\{\psi^{(j;l)}, \zbf^{(j;l)}\}_{l=1}^{N_j}$ are obtained from
$\pi_{\xi_j}(\psi,\zbf|\ybf)$, $j=1,\ldots,k$ to estimate
$m_{\xi}(\ybf)/m_{\xi_1}(\ybf)$ for all $\xi$ using
\eqref{eq:dhatbfsg}. \cite{roy:tan:fleg:2017} provide standard error
estimates of $\hat {\rbf}$ and $\hat{B}_{\xi, \xi_1} (\hat {\rbf})$
that can be used for deciding the appropriate sample sizes
$\tilde{N}_j$'s and $N_j$'s.  This two-stage GIS estimator was used in
\cite{roy:evan:zhu:2016} for EB estimation in the binomial SGLMM with
robit link. More details about this procedure are given in
Appendix~\ref{sec:summ}.
 However, as with the naive IS method,
the variability of~\eqref{eq:ptil} can be high if the importance densities
do not sufficiently ``overlap''. This issue is overcome using reparameterizations.

\subsection{Reparameterized generalized importance sampling estimators}
\label{sec:gis}

It turns out that, under certain circumstances, the GIS estimator
\eqref{eq:dhatbfsg} can be unreliable although it is more efficient
than the naive IS estimator \eqref{eq:bf}. The reason is that the
functions $\zbf \mapsto p[\ybf | \mubf = f_{\nu} (\zbf)]$ and
$\zbf \mapsto p[\ybf | \mubf = f_{\nu'} (\zbf)]$ can be very different
when $\nu \neq \nu'$ \cite[see e.g.][]{chri:2004}. Consequently, the
Monte Carlo sample will be {\it separable} \citep[see][]{geye:1994} if the
points in the skeleton set are not sufficiently close. For example,
suppose $\zbf$ is a sample (generated by a Metropolis-Hastings
algorithm) from the ($\zbf$ marginal) posterior density \eqref{eq:compost} corresponding
to the Poisson SGLMM with the Box-Cox link with exponent $\nu=1$. Thus,
most likely, the sampled $\zbf$ assigns appreciable mass to the
probability $p[\ybf|\mubf = \zbf+1]$. The RL estimator
and~\eqref{eq:dhatbfsg} require that we evaluate
$p[\ybf | \mubf = f_{\nu'} (\zbf)]$ at all other $\nu'$ in the
skeleton set $\Xi$. If $\nu'=0$ this becomes
$p[\ybf| \mubf = \exp(\zbf)]$ so the mean of the Poisson distribution
changes drastically even for moderate values of $\zbf$, and, in
effect, the probability corresponding to $\nu'$ can be numerically
indistinguishable from 0.

To avoid this issue, we consider reparameterizations of the integral
in~\eqref{eq:likgls}. To that end, write the likelihood in~\eqref{eq:likgls}
as an integral with respect to $\mubf$ instead of $\zbf$. Consider the
transformation $h_\nu^{-1}: \zbf \mapsto \mubf$ which is valid only when
the link function is invertible over the whole real line. The Jacobian of
the transformation is $\tilde J_\nu(\mubf) = \prod_{i=1}^n h'_\nu(\mu_i)$.
As in~\eqref{eq:compost}, the corresponding complete posterior density of
$(\psi, \mubf)$ is
\begin{equation}
  \label{eq:postsimu}
\pi_\xi( \psi, \mubf| \ybf) = \frac{p[\ybf, \mubf | \psi, \xi]
\pi(\psi)}{m_{\xi}(\ybf)},
\end{equation}
 based on the augmented joint density
$p[\ybf, \mubf | \psi, \xi] = p[\ybf| \mubf] p[\zbf = h_\nu (\mubf) | \psi,
\xi] \tilde J_\nu(\mubf)$. Note that if we have a Markov chain
$\{\psi^{(i)}, \zbf^{(i)}\}_{i \ge 0}$ with stationary density
$\pi_\xi(\psi, \zbf| \ybf)$ then
$\{\psi^{(i)}, \mubf^{(i)} = f_\nu(\zbf^{(i)})\}_{i \ge 0}$ is a Markov
chain with stationary density $\pi_\xi(\psi, \mubf| \ybf)$ given in \eqref{eq:postsimu}. The advantage
of using the latter is that the estimator~\eqref{eq:dhatbfsg} now becomes
\begin{equation}
  \label{eq:glsbf2}
  \tilde{B}_{\xi, \xi_1} (\tilde{\rbf}) = \sum_{j =1}^k \sum_{l =1}^{N_j}
  \frac{ p[\zbf = h_\nu (\mubf^{(j;l)}) | \psi^{(j;l)}, \xi]
    \tilde J_\nu(\mubf^{(j;l)})}{\sum_{i=1}^k N_i
    p[\zbf = h_{\nu_i} (\mubf^{(j;l)})| \psi^{(j;l)}, \xi_i] \tilde J_{\nu_i}(\mubf^{(j;l)})/
    \tilde{r}_i},
\end{equation}
which, unlike \eqref{eq:dhatbfsg}, does not involve the conditional pmf of
$\ybf$, $p[\ybf |\mubf = f_\nu(\zbf)]$. In \eqref{eq:glsbf2}, we use
$\tilde{r}_i$ to denote RL estimator of $r_i$ based on the transformed MC
samples, i.e., by using
$p[\zbf = h_{\nu_j} (\mubf) | \psi, \xi_j] \tilde J_{\nu_j}(\mubf)$ instead
of $p[\ybf, \zbf|\psi, \xi_j]$, for $j=1,\dots, k$ in \eqref{eq:ptil}.

The use of~\eqref{eq:glsbf2} presents two new challenges. First, it is
valid only when the link function maps onto the whole real line, therefore it
cannot, in general, be used with the Box-Cox link~\eqref{eq:2} or the GEV
link~\eqref{eq:3}. Secondly, computing $h_\nu(\mubf)$ can be slow, which
can add significant computing time when evaluated over many different
values of $\mubf$. Such is the case for the robit link when $\nu < 1$
(see \cite{Rnews:Koenker:2006} and Remark~5 in \cite{AS109}).

More generally, we can use any transformation
$g^{-1}_\nu: \zbf \mapsto \wbf$, not necessarily the link. Here
$\wbf=(w_1,\dots,w_n)$. If chosen appropriately such that
$w_i \approx \mu_i$, it can alleviate the separability problem. For
example, if Box-Cox or GEV link is used for analyzing data, then the
modified versions of the Box-Cox and GEV links introduced in
sections~\ref{sec:bin} and \ref{sec:pois} can be used for constructing this
transformation. In the case of the robit link, it can be the Wallace
transformation~\eqref{eq:6} which is computationally faster. Define the
complete posterior density of $(\psi, \wbf)$,
\begin{equation}
  \label{eq:postsiw}
  \pi_{\xi}(\psi, \wbf| \ybf)=\frac{p[\ybf|\mubf = f_\nu( g_\nu (\wbf))]
  p[\zbf = g_\nu (\wbf) | \psi, \xi] \bar
  J_\nu(\wbf)\pi(\psi)}{m_{\xi}(\ybf)},
\end{equation}
where $\bar J_\nu(\wbf) = \prod_{i=1}^n g_\nu'(w_i)$. The estimator of
the Bayes factors in the general case becomes
\begin{equation}
  \label{eq:glsbf3}
  \bar{B}_{\xi, \xi_1} (\bar \rbf) = \sum_{j =1}^k \sum_{l =1}^{N_j}
  \frac{ p[\ybf|\mubf = f_\nu( g_\nu (\wbf^{(j;l)}))] p[\zbf = g_\nu
    (\wbf^{(j;l)}) | \psi^{(j;l)}, \xi]
    \bar J_\nu(\wbf^{(j;l)})}{\sum_{i=1}^k N_i p[\ybf|\mubf = f_{\nu_i}( g_{\nu_i}
    (\wbf^{(j;l)}))]
    p[\zbf = g_{\nu_i} (\wbf^{(j;l)})| \psi^{(j;l)}, \xi_i] \bar J_{\nu_i}(\wbf^{(j;l)})/
    \bar{r}_i},
\end{equation}
where $\{\psi^{(j;l)}, \wbf^{(j;l)}\}_{l \ge 0}$ is a Markov chain
with stationary density $\pi_{\xi_j}(\psi, \wbf| \ybf)$ given in
\eqref{eq:postsiw}, and $\bar r_i$ is the RL estimator of $r_i$ based on
the samples $\{\psi^{(j;l)}, \wbf^{(j;l)}\}_{l \ge 0}$. As before, if
we have a Markov chain $\{\psi^{(i)}, \zbf^{(i)}\}_{i \ge 0}$ with
stationary density $\pi_\xi(\psi, \zbf| \ybf)$ then
$\{\psi^{(i)}, \wbf^{(i)} = g_\nu^{-1}(\zbf^{(i)})\}_{i \ge 0}$ is a
Markov chain with stationary density $\pi_\xi(\psi, \wbf| \ybf)$.
Unlike \eqref{eq:glsbf2}, the expression of \eqref{eq:glsbf3} is not
free of the pmf of $\ybf$, but, as we show through examples in
Section~\ref{sec:exam}, \eqref{eq:glsbf3} can lead to huge gains in
computational efficiency over \eqref{eq:glsbf2} without sacrificing
accuracy. Note that the
GIS estimators~\eqref{eq:dhatbfsg} and~\eqref{eq:glsbf2} are special
cases of~\eqref{eq:glsbf3} with $g_\nu$ being the identity function
and $f_\nu^{-1}$ respectively. In \eqref{eq:glsbf3} the RL estimator
$\bar {\rbf}$ is obtained by using
$p[\ybf|\mubf = f_{\nu_j}( g_{\nu_j} (\wbf))] p[\zbf = g_{\nu_j}
(\wbf) | \psi, \xi_j] \bar J_{\nu_j}(\wbf)$ instead of
$p[\ybf, \zbf|\psi, \xi_j],$ for $j=1,\dots, k$ in \eqref{eq:ptil}.

Note that the function $g_\nu$ can be different for each component of
the vector $\zbf$, so we can apply a different transformation to each
component. One example where we want to do that is the case of the
Poisson SGLMM with the Box-Cox link where some $y_i$'s are strictly
positive, and some other $y_i$'s are equal to zero.  As we have already
explained, in the latter case the Box-Cox link does not map onto the
real line so the modified Box-Cox transformation should be used. In
case of $y_i > 0$, we must have $\mu_i > 0$, but this can fail if $w_i$
is simulated conditional on $\nu=\nu_1$ and
$\mu_i = f_\nu ( g_\nu (w_i))$ is evaluated at $\nu = \nu_2 > \nu_1$
when $g_\nu$ is the modified Box-Cox function. Therefore, we let
$g_\nu$ be the Box-Cox transformation when $y_i>0$ and the modified
transformation when $y_i=0$.

The estimator~\eqref{eq:glsbf3} can be further improved by
the use of control variates \citep{owen:zhou:2000}. The use of control variates in the context
of GIS estimation was discussed in \cite{doss:2010}. Below we use
control variates to improve the reparameterized estimator \eqref{eq:glsbf3}. Let
$a_i = N_i/N$,
\begin{equation*}
  q_\xi(\psi, \wbf) = p[\ybf|\mubf = f_\nu( g_\nu (\wbf))] p[\zbf = g_\nu
  (\wbf) | \psi, \xi]  \bar J_\nu(\wbf),
\end{equation*}
and define
\begin{equation}
\label{eq:cvy}
  Y(\psi, \wbf) = \frac{q_\xi(\psi, \wbf)}{\sum_{i=1}^k a_i
    q_{\xi_i}(\psi, \wbf)/ {r}_i},
\end{equation}
and for $j=2,\ldots,k$,
\begin{equation}
\label{eq:cvx}
  X_j(\psi, \wbf) = \frac{q_{\xi_j}(\psi, \wbf)/{r}_j - q_{\xi_1}(\psi, \wbf)}
  {\sum_{i=1}^k a_i q_{\xi_i}(\psi, \wbf)/ {r}_i}.
\end{equation}
Note that $\E Y(\psi, \wbf) = B_{\xi,\xi_1}$ and $\E X_j(\psi, \wbf) = 0$ where
the expectation is taken with respect to the mixture density
\begin{equation}
\label{eq:postmix}
  \pi_\text{mix}(\psi, \wbf | \ybf) = \sum_{i=1}^k a_i
  \pi_{\xi_i}(\psi, \wbf|\ybf) .
\end{equation}
Then, for any $\alpha = (\alpha_2,\ldots,\alpha_k)$,
\begin{equation}
  \label{eq:7}
  \hat{I}_\alpha = \frac{1}{N} \sum_{j=1}^k \sum_{l=1}^{N_j}
  \left\{ Y(\psi^{(j;l)}, \wbf^{(j;l)}) - \sum_{i=2}^k \alpha_i X_i(\psi^{(j;l)}, \wbf^{(j;l)}) \right\} ,
\end{equation}
is an unbiased estimator of $B_{\xi,\xi_1}$ where the samples are
obtained from the density \eqref{eq:postmix}. In the case $\alpha=0$,
$\hat{I}_\alpha$ reduces to $\bar{B}_{\xi,\xi_1}(\rbf)$, but
\cite{owen:zhou:2000} argued that an optimal choice for $\alpha$ is
the one that minimizes the variance of~\eqref{eq:7}, in which case
$\hat{I}_\alpha$ corresponds to the estimate of the intercept term in
the least squares regression of $Y(\psi^{(j;l)}, \wbf^{(j;l)})$
against $X_i(\psi^{(j;l)}, \wbf^{(j;l)}), i=2,\dots,k$. In practice,
$r_i$ is replaced by its reverse logistic regression estimate,
$\bar{r}_i$, in \eqref{eq:cvy} and \eqref{eq:cvx} before computing $\hat{I}_\alpha$.

\subsection{Derivation of skeleton points}
\label{sec:skel}

In this section we describe a method of choosing the multiple importance
densities corresponding to the mixture distribution used in the
GIS estimator~\eqref{eq:dhatbfsg} and its
derivatives based on transformed samples. This boils down to choosing the
skeleton set $\Xi$.

Because $B_{\xi,\xi_1} \propto m_\xi(\ybf)$, the skeleton set is derived by
approximating the integral in~\eqref{eq:marg} using integrated, nested
Laplace approximations. The approximation can be used to get preliminary
estimates of $m_\xi(\ybf)$ and thus of $\hat{\xi}$. Consequently, we derive
a range of ``good'' values for skeleton points. The first step is to use
Laplace approximation to approximate the marginal likelihood
$L_\xi(\sigma^2|\ybf)$ for given $\sigma^2$ and the second step is to
integrate out $\sigma^2$ numerically, so the first step is nested within
the second step. This method is presented below with further details in
Appendix~\ref{sec:apdx:lapl-appr}.

First consider the integral in~\eqref{eq:marg}. Under the Gaussian prior
assumption for $\beta$ we can derive the likelihood for $\sigma^2$ for
given $\xi$ as,
\begin{equation*}
  L_\xi(\sigma^2|\ybf) = \int_{\mathcal{R}^n} p[\ybf, \zbf | \sigma^2, \xi] \ud \zbf,
\end{equation*}
where $p[\ybf, \zbf | \sigma^2, \xi] = p[\ybf|\mubf = f_\nu(\zbf)] p[\zbf|\sigma^2,\xi]$,
with $p[\zbf|\sigma^2,\xi] = \int_{\mathcal{R}^p} p[\zbf|\beta, \sigma^2,\xi] \pi(\beta) \ud\beta$ being a Gaussian density.

Let
\begin{align}
\label{eq:htil}
  \tilde{\zbf}_\xi(\sigma^2) &= \argmax_\zbf p[\ybf,\zbf|\sigma^2,\xi] , \nonumber\\
  \tilde{H}_\xi(\sigma^2) &= -\frac{\partial^2}{\partial \zbf \partial \zbf^\T} \log
  p[\ybf,\zbf|\sigma^2,\xi] \bigr|_{\zbf = \tilde{\zbf}_\xi(\sigma^2)} .
\end{align}
Then, by Laplace approximation \citep{barndorff1989asymptotic},
\begin{equation*}
  L_\xi(\sigma^2|\ybf) \approx p[\ybf|\mubf=f_\nu(\tilde\zbf_\xi(\sigma^2))]
  p[\zbf=\tilde\zbf_\xi(\sigma^2)|\sigma^2,\xi]
  \left|\frac{1}{2\pi} \tilde{H}_\xi(\sigma^2) \right|^{-\half} ,
\end{equation*}
for any given $\sigma^2$. Using this result in~\eqref{eq:marg} we have
\begin{equation}
  \label{eq:LAmarg}
  m_\xi(\ybf) = \int_0^\infty L_\xi(\sigma^2|\ybf) \pi(\sigma^2) \ud \sigma^2 \approx
  \int_0^\infty p[\ybf|\mubf=f_\nu(\tilde\zbf_\xi(\sigma^2))]
  p[\zbf=\tilde\zbf_\xi(\sigma^2)|\sigma^2,\xi]
  \left|\frac{1}{2\pi} \tilde{H}_\xi(\sigma^2) \right|^{-\half} \pi(\sigma^2)
  \ud \sigma^2 .
\end{equation}
The integration in the right-hand side of~\eqref{eq:LAmarg} is done
numerically using the trapezoid rule in a range of values of $\sigma^2$ where
the integrand has significant mass.

Let $\tilde{m}_{{\xi}}(\ybf)$ denote the approximation in~\eqref{eq:LAmarg}.
To derive a sensible region for the parameter $\xi$, let $\tilde{\xi}$ denote
the maximizer of $\tilde{m}_{{\xi}}(\ybf)$ and let $\tilde{m}_{\tilde{\xi}}(\ybf)$
denote its maximum value. Suppose $\xi$ consists of $d$ components. For
each component $j$, we obtain an interval $(\xi_j^L, \xi_j^U)$ such that
when $\xi_j \in (\xi_j^L, \xi_j^U)$ and the remaining components are equal
to the corresponding components in $\tilde{\xi}$,
$\tilde{m}_{{\xi}}(\ybf)$ is no less than
$\alpha \tilde{m}_{\tilde{\xi}}(\ybf)$ for a predetermined factor
$\alpha \in (0,1)$. Each interval is then discretized to a set of
$T$ equispaced points
$\{\xi_{j}^1 = \xi_j^L, \xi_j^2, \ldots, \xi_j^T = \xi_j^U\}$ and the
discrete points are crossed to create a finite grid of points
$\{\xi_1^1, \xi_1^2, \ldots, \xi_1^T\} \times \ldots \times \{\xi_d^1,
\xi_d^2, \ldots, \xi_d^T\}$. The points $\xi$ in this grid where
$\tilde{m}_{{\xi}}(\ybf) < \alpha \tilde{m}_{\tilde{\xi}}(\ybf)$ are
discarded and the remaining points, $\Xi$, define the skeleton set. If
the number of points in $\Xi$ is deemed large for the available
computational resources, then $\alpha$ is increased accordingly.

\subsection{Model choice and weighting}
\label{sec:msel}

In practice, information about the true underlying model is
limited. Using parametric links can make inference more robust but
this still assumes a specific parametric form for the link function
and correlation function. So far we have discussed how to choose
between models with the same parametric link and correlation
function. In this section we discuss choosing between different links,
and correlation families.

Suppose there are $R$ candidate models each of which specify a link and
a correlation function, denoted by $\Mcal_r(\xi_r)$, with
associated parameters $\xi_r$, for $r = 1,\ldots,R$. We write $p[\ybf,\wbf|\psi;\Mcal_r(\xi_r)]
$ for the joint density of $\ybf$ and $\wbf$ under model $\Mcal_r(\xi_r)$. Then, the
corresponding marginal density for the data is
\begin{equation*}
  m_{r,\xi_r}(\ybf) = \int_{\mathcal{R}^p \times \mathcal{R}_+}\int_{\mathcal{R}^n} p[\ybf,\wbf|\psi;\Mcal_r(\xi_r)] \pi(\psi) d\wbf \ud\psi .
\end{equation*}
Although the marginal density of $\ybf$ remains the same
whether integrated with respect to $\wbf$ or $\zbf$, we use the joint
density of $\ybf$ and $\wbf$ because, later in this section, the RL
estimation is used with reparameterized samples. Note that $\wbf$ is
any transformed version of $\zbf$, so it can be $\zbf$ if we
let $g_\nu$ to be the identity function.

A general measure of model comparison is the AIC which is defined as
\begin{equation*}
  \mathrm{AIC}_r = -2\log m_{r,\hat{\xi}_r}(\ybf) + 2 d_r,
\end{equation*}
where $\hat{\xi}_r$ is the EB estimate of $\xi_r$ and $d_r$ is the
number of parameters in $\xi_r$. A model with lower AIC value would be
preferred, although it can also be used for model weighting in the
spirit of \cite{buckland1997model}, an approach we come to at the end of
this section. The AIC formula is not
straightforward to apply because we don't know the value of
$m_{r,\hat\xi_r}(\ybf)$. In Section~\ref{sec:gis} we have discussed
how the RL method is used to estimate ratios $m_{r,\xi_r}(\ybf)/m_{r,\xi_{r,1}}(\ybf)$
for models having the same functional forms for the link function and
the spatial covariance functions, i.e. within $\Mcal_r$. Although, in
principle, the GIS methods developed in Section~\ref{sec:gis} may be
used to estimate Bayes factors across different models $\Mcal_r$'s, it
is computationally demanding as large number of skeleton points with
several combinations of $\xi_r$ values from these models need to be used
for accurate estimation of Bayes factors.
Instead, we consider minimizing
\begin{equation*}
  \mathrm{AIC}^*_r = -2\log
  \frac{m_{r,\hat{\xi}_r}(\ybf)}{m_{1,\hat{\xi}_1}(\ybf)} + 2 d_r,
\end{equation*}
and apply
the RL method to estimate 
the ratios ${m_{r,\hat{\xi}_r}(\ybf)}/{m_{1,\hat{\xi}_1}(\ybf)}$ for
$r=1,\ldots,R$. 

Suppose $\{\psi^{(l,r)}, \wbf^{(l,r)}\}_{l=1}^{L_r}$ is a Harris ergodic
Markov chain with stationary density 
$\pi(\psi, \wbf|\ybf;\Mcal_r(\hat{\xi}_r))$ corresponding to the model
$\Mcal_r(\hat{\xi}_r)$, $r=1,\ldots,R$. Let
\begin{equation*}
  C_r = m_{r,\hat{\xi}_r}(\ybf)/ m_{1,\hat{\xi}_1}(\ybf),
\end{equation*}
and
\begin{equation*}
  \eta_r = -\log C_r + \log \frac{L_r}{L},
\end{equation*}
where $L = \sum_r L_r$ and
\begin{equation*}
  P_r(\psi, \wbf;\eta) = \frac{
    p[\ybf,\wbf|\psi;\Mcal_r(\hat{\xi}_r)] e^{\eta_r}}{\sum_{s=1}^R
    p[\ybf,\wbf|\psi;\Mcal_s(\hat{\xi}_s)] e^{\eta_s}} ,
\end{equation*}
where $\eta = (\eta_1,\ldots,\eta_R)$.
Estimation of $\eta$ is possible up to an additive constant using the
samples $\{\psi^{(l,r)}, \wbf^{(l,r)}\}_{l=1}^{L_r}, r=1,\dots,R$ by maximizing the
quasi log-likelihood
\begin{equation}
  \label{eq:5}
  \mathcal{L}(\eta) = \sum_{r=1}^R \sum_{l=1}^{L_r} \log
  P_r(\psi^{(l,r)}, \wbf^{(l,r)};\eta) .
\end{equation}
Let $\hat{\eta}$ denote the maximizer of~\eqref{eq:5} subject to the
constraint that $\sum \hat{\eta}_r = 0$, and let $\hat{C}_r =
\frac{L_r}{L} e^{-\hat{\eta}_r}$ be the
corresponding estimate of $C_r$. Then, an estimate of
$\mathrm{AIC}^*_r$ is
\begin{equation*}
\label{eq:aichat}
  \widehat{\mathrm{AIC}}^*_r = -2\log \hat{C}_r + 2 d_r,
\end{equation*}
and we choose the model with the smallest $\widehat{\mathrm{AIC}}^*_r$ value.

Instead of choosing a single model, \cite{buckland1997model} argue for an
ensemble modeling approach where a model weight is calculated from the AIC
values. In our case, we define the weight for the $r$th model to be
\begin{equation}
  \label{eq:12}
\mbox{Weight}_r = \frac{\exp(-\widehat{\mathrm{AIC}}^*_r/2)}{\sum_{t=1}^R \exp(-\widehat{\mathrm{AIC}}^*_t/2)}.
\end{equation}
Let $\hat\mu_r(s)$ denote the estimated mean response at spatial location
$s \in \mathbb{S}$ using the model $\mathcal{M}_r(\hat{\xi}_r)$, $r \in \{1,\ldots,R\}$.
Then, an ensemble estimate of the mean response at that location is given
by
\begin{equation}
  \label{eq:11}
  \hat\mu(s) = \sum_{r=1}^R \mbox{Weight}_r \times \hat\mu_r(s).
\end{equation}

\section{Simulations}
\label{sec:exam}
In this section using simulation examples, we demonstrate the
advantages of using transformed samples in GIS estimation. Simulation
studies are also used to exhibit the performance of the proposed model
selection criterion in choosing the true link functions and the spatial
covariance structures.
All analyses in this paper are performed using the R
package geoBayes \citep{r:geob}.

\subsection{Comparison with the untransformed estimator for the
  binomial-robit model}
\label{sec:binomial-robit-model}

The purpose of this section is to demonstrate that the GIS estimator
based on the untransformed samples (method of
\cite{roy:evan:zhu:2016}) can be biased for estimating $\xi$ when the
skeleton set is not dense enough. The reason for this bias is the
little ``overlap'' among the importance densities. On the other hand,
the reparameterized version~\eqref{eq:glsbf2}, although unbiased,
is much slower because computing the robit link function for degrees
of freedom $\nu < 1$ is slow.  Instead, \eqref{eq:glsbf3} with
the Wallace link reparameterization~\eqref{eq:6} provides unbiased
estimates, and is also faster than~\eqref{eq:glsbf2}.

We consider the spatial domain $\Sbb = [0,1] \times [0,1]$ and
randomly select $n=100$ locations $s \in \Sbb$ to sample from. The
spatial random field $z(s)$ is assumed to have exponential correlation
structure with unknown spatial range parameter $\phi = 0.5$ and variance
$\sigma^2 = 1$. The mean of the random field is taken to be constant $\beta=-1$.

In this section the response variable is conditionally binomially
distributed given the value of the spatial field with number of trials
$t = 100$ at each sampling location and the probability of success at
location $s$, $\mu(s)$, is given by
\begin{equation*}
  \mu(s) = G_\nu (z(s)),
\end{equation*}
where $G_\nu (\cdot)$, as defined in
section~\ref{sec:sglm}, is the cdf of the standard Student's $t$
distribution with $\nu$ degrees of freedom. Here we take $\nu = 0.5$
for simulating the data. The parameters $\beta$ and $\sigma^2$ are
assigned normal and scaled-inverse-chi-square priors as discussed in
Section~\ref{sec:eb} with hyperparameter values $m_b = 0$,
$V_b = 100$, $n_\sigma = 1$, and $a_\sigma = 1$. The link parameter
$\nu$ and spatial range parameter $\phi$ are then treated as unknown
and are estimated using the EB procedure of section~\ref{sec:eb}.  The
skeleton set for the parameters $\xi = (\nu,\phi)$ is set to
\begin{equation*}
  \Xi = \{0.4, 1, 3, 7, 14\} \times \{0.25, 0.7, 1\} .
\end{equation*}

For fixed $\xi \in \Xi$, we sample from
$\pi_\xi(\beta, \sigma^2, \zbf| \ybf)$, the complete posterior density
of $\beta$, $\sigma^2$ and the random field $\zbf$ conditional on the
observed data $\ybf$. For each $\xi \in \Xi$, we obtain a Markov chain
sample of size 1000 after a burn-in of 300 samples and thinning of
5. From these samples, 800 samples were used to obtain RL estimate
$\hat {\rbf}$, and the remaining 200 samples were used to calculate the GIS
estimator $B_{\xi,\xi_1}(\hat {\rbf})$. We computed the three GIS
estimators given in~\eqref{eq:dhatbfsg},~\eqref{eq:glsbf2},
and~\eqref{eq:glsbf3}. Once $\hat \xi$ is estimated using these GIS
estimators, posterior means of the parameters $(\beta, \sigma^2)$ are
estimated based on Markov chain samples of length $1000$ after a
burn-in of 300 samples and thinning of 5 from the posterior density
$\pi_{\hat \xi}(\psi| \ybf)$.

\begin{figure}
  \centering
  \includegraphics[width=\linewidth]{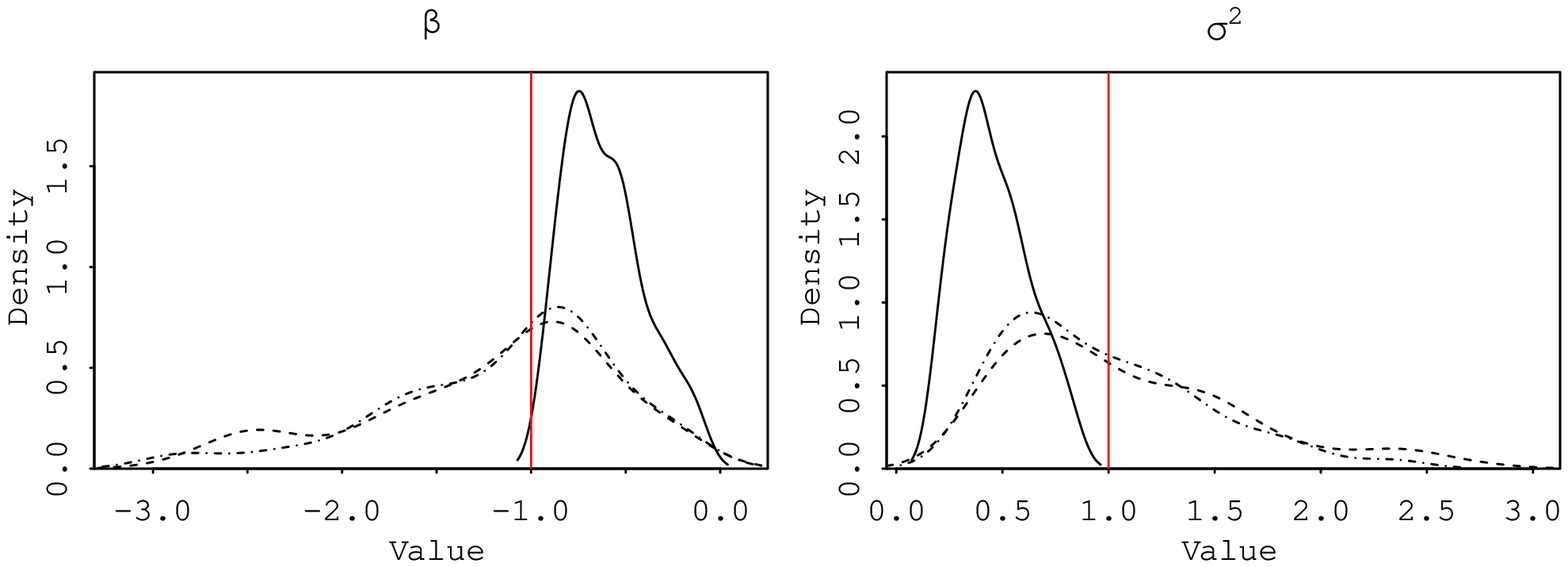}\\
  \includegraphics[width=\linewidth]{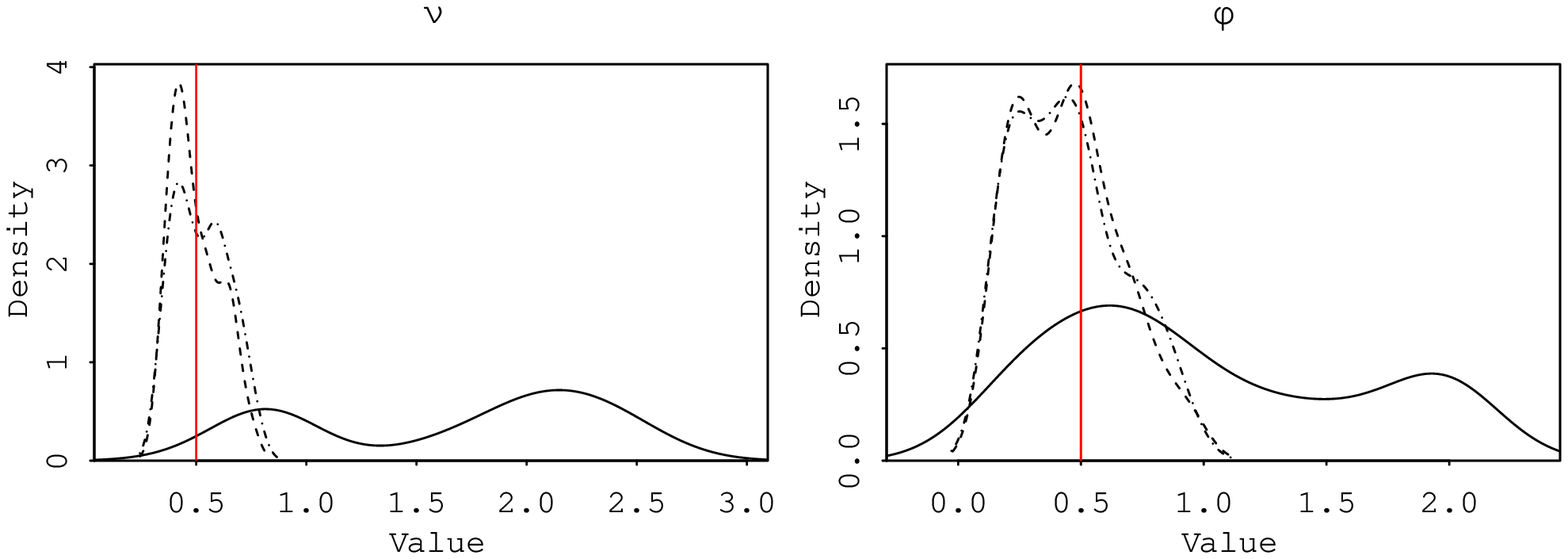}
  \caption{Distribution of parameter estimates over all simulations for the
    binomial robit model using different transformations based GIS: None
    (solid); Link (dashed); Wallace (dashed-dotted). The true parameter
    value is shown by a vertical line.}
  \label{fig:bin_robit1}
\end{figure}

We performed 100 simulations where the sampling locations remained the
same but the spatial random field and the observations were
different. In Figure~\ref{fig:bin_robit1} we show the kernel density
of the parameter estimates using each of the three methods described
in this paper. It is clear that the GIS estimator \eqref{eq:dhatbfsg} with
untransformed samples can lead to incorrect inference while the two
methods based on the transformed samples do not exhibit such bias. The
exact biases are shown in Table~\ref{tab:bin_robit1} along with the mean
square difference from the true value of the spatial field, which also
shows that the untransformed estimator has the worst performance. On
the other hand, as shown in Table~\ref{tab:bin_robit1}, using the link
transformation~\eqref{eq:glsbf2} can be slow for $\nu < 1$. The
alternative transformation method~\eqref{eq:glsbf3} is much faster
although perform similarly as~\eqref{eq:glsbf2}.

\begin{table}
  \centering
  \begin{tabular}{lccccccc}
    \toprule
    & Bias($\nu$) & Bias($\phi$)
    & Bias($\beta$) & Bias($\sigma^2$) & MSE($\zbf$) & Time 1st stage
    & Time 2nd stage \\
    \midrule
    None    & 1.17 &    0.50 &    0.39 & $-0.55$ & 0.50 & 28 & 61 \\
    Link    & 0.00 & $-0.05$ & $-0.22$ &    0.06 & 0.29 & 96 &315 \\
    Wallace & 0.02 & $-0.05$ & $-0.16$ & $-0.05$ & 0.27 & 38 &114 \\
    \bottomrule
  \end{tabular}
  \caption{Bias of the empirical Bayes estimates for $\nu$ and $\phi$, bias
    of the posterior estimates of $\beta$ and $\sigma^2$, mean
    square error of the posterior for $\zbf$, and computing times in
    seconds for the first and second stage per iteration for the simulation
    example of Section~\ref{sec:binomial-robit-model}.}
  \label{tab:bin_robit1}
\end{table}

\subsection{Model selection}
\label{sec:model-selection}

In this section we demonstrate that the model selection criterion of
Section~\ref{sec:msel} chooses the correct model for the link and
correlation functions. We randomly choose $n=100$ locations to sample from
within the spatial domain $\Sbb = [0,1] \times [0,1]$. Samples are taken
from two models. Model M1 is the binomial SGLMM with robit link and
exponential correlation, and model M2 is the binomial SGLMM with modified
GEV link and spherical correlation. For both models the spatial random
field was sampled with spatial range $\phi = 0.5$, relative nugget
$\omega = 0$ (assumed known), variance $\sigma^2 = 1$, and mean
$\beta = 0.5$. Conditioned on the value of the spatial field, the
observation at the $i$th location was sampled from the binomial
distribution with number of trials $t_i=100$ for all $i$ and probability
of success $F_\nu(z_i)$ where $F_\nu(\cdot)$ is the cdf
of the standard Student's $t$ distribution with degrees of freedom $\nu = 0.5$ for
model M1, and the function in~\eqref{eq:4} with $\nu=0$ for model M2. Our models
are completed by assuming a scaled-inverse-chi-squared prior for $\sigma^2$
with degrees of freedom 1 and scale 1, and a conditional normal prior for
$\beta$ given $\sigma^2$ with mean 0 and variance $10 \sigma^2$.

From
each model, 100 different data sets were simulated from the same
100 locations but with different spatial random field each time. For each simulated data set, we fit nine different models by assuming
three different link functions: robit, probit, and modified GEV,
combined with three different correlation functions: exponential,
Gaussian, spherical.  The skeleton set for $\xi = (\nu,\phi)$, $\Xi$,
for each model corresponds to $\Xi = \Xi_\nu \times \Xi_\phi$ where
$\Xi_\nu = \{0.4, 1, 3, 7, 15\}$ for the robit link, $\Xi_\nu$ is the
null set for the probit link, $\Xi_\nu = \{0, 0.5, 1\}$ for the
modified GEV link, and $\Xi_\phi = \{0.3, 0.7, 1.1\}$ for all three
correlation functions considered. We fit each model by first
estimating $\xi$ by $\hat\xi$ from maximizing the reparameterized
estimator~\eqref{eq:glsbf2}, and then calculating each model's weight
using~\eqref{eq:12}. The estimation of $\xi$ is done by generating
MCMC samples from the posterior distribution of $(\psi,\zbf)$
conditioned on a value of $\xi$ in the skeleton set. The size of MCMC
samples, after a burn in of 300 samples, is $10^4/k$ rounded down
where $k$ is the size of the skeleton set. From these samples
approximately 80\% is used for stage~1, and the remaining samples are
used for stage~2 of the procedure described in Appendix~B. To estimate
the model weights we use 1000 MCMC samples, after a burn in of 300
samples, from the posterior density of $(\psi,\zbf)$ conditioned on
the value of $\hat\xi$ from each model.

The average weight for each model over the 100 different simulated datasets
is given in Table~\ref{tab:model_weight}. In both cases the correct model
has the highest average weight. The results show that the
proposed approach is very good in selecting the correct link function among
those considered. Due to the similarities between the exponential and
spherical correlations \citep[][Sec~2.10]{stei:1999} the second-best model
chooses either of these when the true model is the other, however, the
Gaussian model, which is not close to the true model, is not favored by our
criterion.

\begin{table}
  \centering
  \begin{tabular}{lccccccccc}
    \toprule
Link    & robit & probit & mGEV & robit & probit &
mGEV & robit & probit & mGEV \\
Correlation & exp & exp & exp & Gau & Gau
& Gau & spher & spher & spher \\
    \midrule
M1 (robit, exp)  & 0.36 & 0.02 & 0.08 & 0.03 & 0.06 & 0.03 & 0.32 & 0.02 & 0.08 \\
M2 (mGEV, spher) & 0.05 & 0.07 & 0.32 & 0.00 & 0.00 & 0.00 & 0.07 & 0.08 & 0.41 \\
\bottomrule
  \end{tabular}
  \caption{Average model weight for each model given in the columns using
    data generated from the model given in the rows.}
  \label{tab:model_weight}
\end{table}

\section{Examples with real data analysis}
\label{sec:examples}
This section illustrates the proposed link functions, reparameterized GIS estimators
and the EB methodology using binomial and Poisson SGLMMs fitted to
analyze two real spatial data sets.

\subsection{Analysis of radionuclide concentrations on the Rongelap
  island}
\label{sec:rongelap}

The dataset consists of the measurements of $\gamma$-ray counts $y_i$
observed during $t_i$ seconds at $i$th coordinate on the Rongelap
island, $i=1,\ldots,n$, $n=157$. This data set was analyzed by
\cite{digg:tawn:moye:1998} and \cite{chri:2004}, among others, using a
Poisson spatial model. Using likelihood analysis, \cite{chri:2004}
found that the Box-Cox link with $\nu = 0.84$ was more appropriate for
these data if an exponential correlation is used. Here we demonstrate
the application of EB methodology developed in section~\ref{sec:eb} on
this example.

Our model consists of a Poisson SGLMM with the modified Box-Cox link
function for the $\gamma$-ray counts. For the spatial Gaussian random field
we fit a constant mean $\beta$ and covariance consisting of a partial sill
parameter $\sigma^2$, a relative nugget term $\omega$, and a correlation
function parameterized by $\theta$, which is yet to be determined. The 
parameters $\beta$ and $\sigma^2$ are assigned the conditional normal and
scaled-inverse-chi-square priors of
Section~\ref{sec:eb} respectively with $m_b=0$, $V_b=100$, $a_\sigma =1$,
and $n_\sigma=1$. In
addition, the parameters $\xi = (\nu,\omega,\theta)$ are also unknown and
are estimated by the EB estimate $\hat{\xi}$ from maximizing~\eqref{eq:glsbf2}.

We consider four different models for the correlation function: Mat\'ern,
exponential-power, spherical, and exponential. All families contain a spatial range
parameter $\phi$ while the first two contain an additional parameter
$\kappa$.

\begin{figure}
  \centering
  \includegraphics[width=\linewidth]{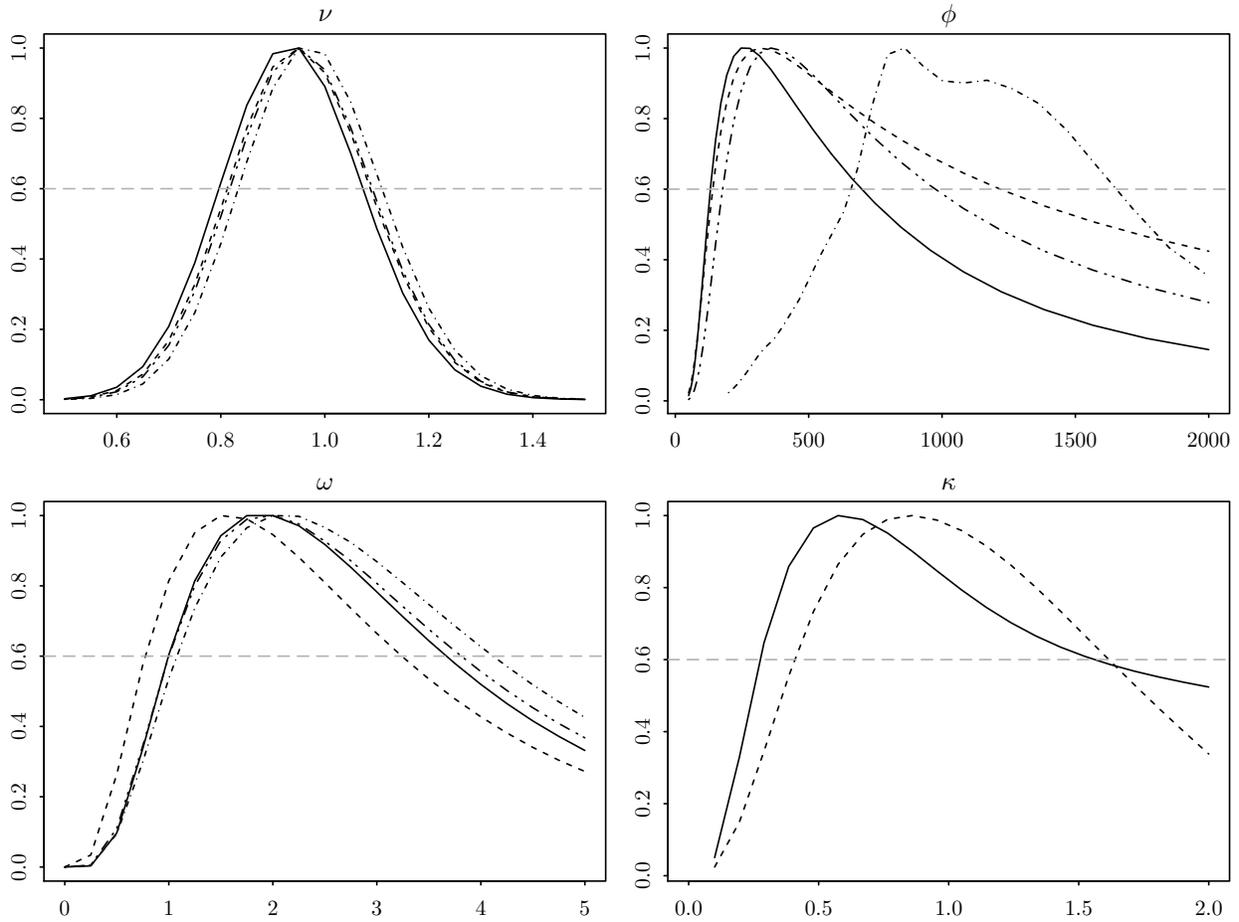}
  \caption{Approximate likelihood computed for a range of parameter values
    for the Rongelap example. The parameter on the horizontal axis varies
    while the other parameters remain fixed at their estimates
    $\tilde{\psi}$. A narrower range is then considered such that the value
    of the likelihood is at least 60\% from its maximum. The four models
    for the correlation function considered are: Mat\'ern (solid);
    exponential-power (dashed); spherical (dashed-dotted); exponential (dashed-dotted-dotted).}
  \label{fig:rongelap_LA}
\end{figure}

At the first stage, we seek a set of skeleton points for the computation of
the GIS estimators. For this we maximize the approximate marginal likelihood
as discussed in Section~\ref{sec:skel}. The maximizer, $\tilde{\xi}$, for
each model can be seen in Figure~\ref{fig:rongelap_LA}. Next we explore the
likelihood for a range of values of $\xi$ around $\tilde{\xi}$. Initially
we fix all but one of the components of $\xi$ at $\tilde{\xi}$ and vary the
other one widely. This allows us to compute the approximate likelihood
quickly for a wide range of each parameter. Then we focus on a narrower
range of the parameters where the marginal likelihood value is at least 60\% from
its maximum (see Figure~\ref{fig:rongelap_LA}). We choose $T=3$ with the
notation of Section~\ref{sec:skel} and evaluate the approximate likelihood
at each combination of parameter values in this narrower range, again
discarding any combinations whose approximate likelihood value falls below
the 60\% threshold. The remaining combinations were used as skeleton points
for the GIS estimators. After this procedure we were left with
12, 10, 5, and 4 skeleton points for the four models respectively, listed
in Appendix~\ref{sec:list-skeleton-points}, Table~\ref{tab:rongelap_skel}. (Note that
the third and fourth models have one less parameter.)

For each set of parameters $\xi$ in the skeleton set, we draw MCMC
samples from the posterior density
$\pi_\xi(\beta, \sigma^2, \zbf |\ybf)$ of the parameters $\beta$,
$\sigma^2$ and the spatial field $\zbf$. The MCMC was run with burn-in
300, while retaining a sample of size $N/k$ (rounded down) where $N=50000$ and $k$ is the number of skeleton points
for each model given in the previous paragraph. Thus the total number of
samples used in the procedure of Appendix~\ref{sec:summ} for each model is about the same. We use approximately 80\% of the
samples for Stage~1 and the remaining 20\% of the samples
for Stage~2. The estimates $\hat{\xi}$ for each model are shown in
Table~\ref{tab:ronge_est}. We also provide standard errors for the EB estimate
$\hat{\xi}$ obtained by the method described in Appendix~\ref{sec:apdx:sterr}.

Subsequently, we fix the parameters $\xi$ at $\hat{\xi}$, and take
a new MCMC sample with burn-in 300, and size 5000. The new sample
is used to estimate the mean parameter $\beta$, the partial sill parameter
$\sigma^2$, and predict the spatial field. Examination of
the posterior samples showed no significant autocorrelations.  The posterior means for the two
parameters are also shown in Table~\ref{tab:ronge_est}. The batch means estimates
of standard errors for the posterior mean estimates are also provided.

Using the new MCMC samples, we also compute the Bayes factors for the three
models relative to the Mat\'ern model as discussed in
Section~\ref{sec:msel}. The estimates of the Bayes factors are shown in
Table~\ref{tab:ronge_est}. It can be seen that the four models have about equal
Bayes factors. The exponential and spherical models
have one fewer parameter, so they are preferable. The estimate of $\nu$ is
slightly higher than \pcite{chri:2004} estimate ($\nu = 0.84$), and
significantly different from the log link ($\nu=0$) used in
\cite{digg:tawn:moye:1998}. We also provide the weight of each model as
given by~\eqref{eq:12}, and use that to calculate ensemble average
estimates of the predicted rate according to~\eqref{eq:11}.

\begin{table}
  \centering
  \begin{tabular}{llllllllll}
    \hline
Model & $\beta$& $\sigma^2$& $\nu$ & $\phi$ & $\omega$ & $\kappa$ & log BF
    & $|\xi|$ & Weight\\
    \hline
    Mat\'ern     & 5.288   & 2.083   & 0.963   & 324 & 2.211& 0.637& 0 & 4 & 0.136 \\
                 & (0.502) & (0.239) & (0.146) & (420) & (1.847) & (0.985) &   &   & \\
       Exp-power & 5.856   & 2.134   & 0.966   & 393 & 2.178& 1.096 & $-0.007$  &4 & 0.135 \\
                 & (0.500) & (0.247) & (0.146) & (336) & (1.957) & (0.917) &   &   & \\
    Spherical    & 5.955   & 1.959   & 0.978   &1170& 2.598 &      & $-0.020$ & 3    &0.363 \\
                 & (0.525) & (0.220) & (0.141) & (332) & (1.810) & &   &   & \\
    Exponential  & 5.780   & 2.129   & 0.957   & 384 & 2.065& & $-0.014$ & 3 & 0.365 \\
                 & (0.501) & (0.244) & (0.145) & (324) & (1.501) &  &   &   & \\
    \hline
  \end{tabular}
  \caption{Parameter estimates with standard errors and log Bayes factor
    relative to the Mat\'ern model for the Rongelap example for each
    model. The standard errors estimates are provided in parentheses. The
    size of $\xi$ is denoted by $|\xi|$.}
  \label{tab:ronge_est}
\end{table}

\begin{figure}
  \centering
  \includegraphics[width=\linewidth]{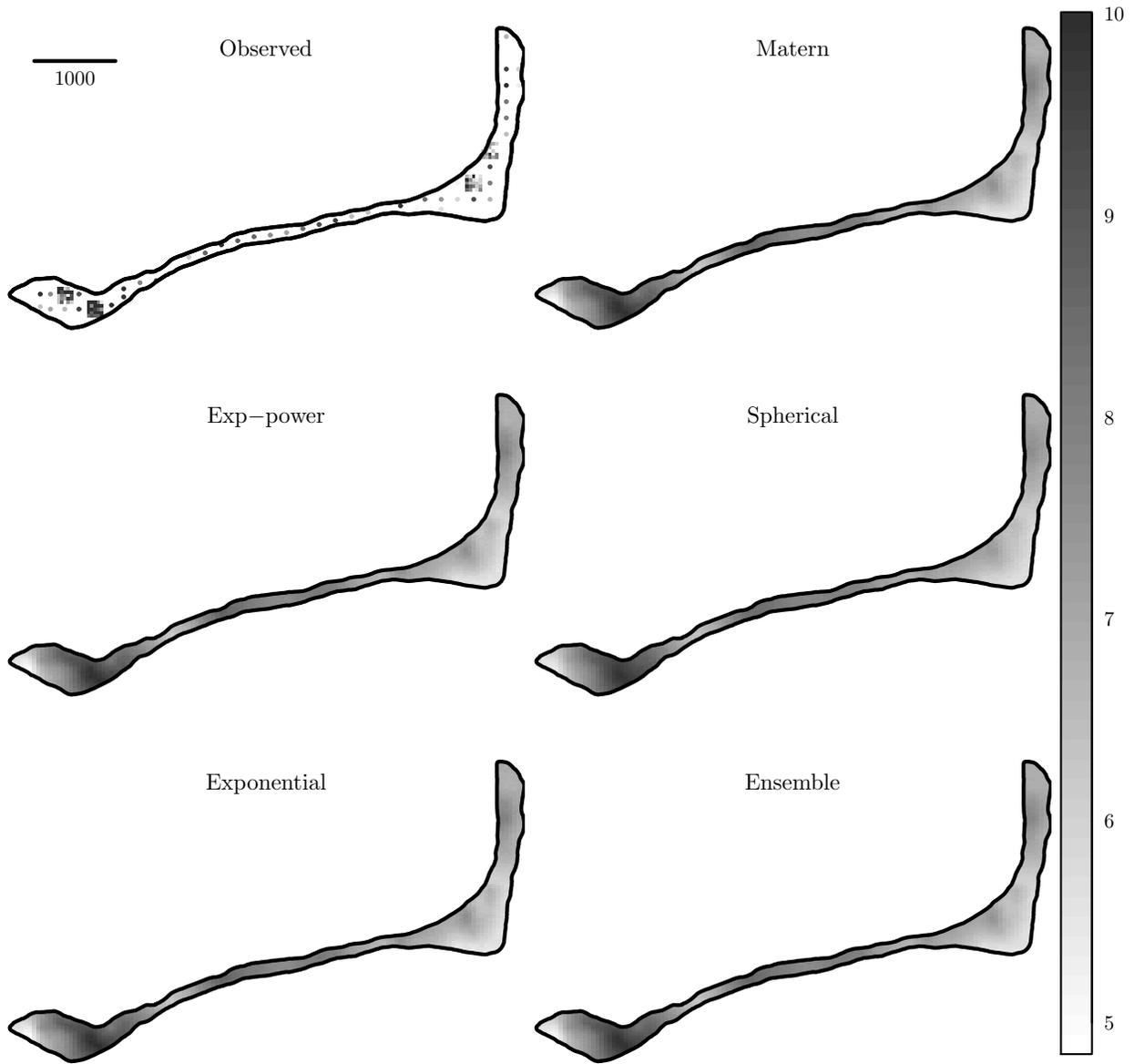}
  \caption{Observed count per unit time and prediction of the Poisson rate (per unit time) for the Rongelap example under
    four different models and ensemble prediction.}
  \label{fig:rongelap_pred}
\end{figure}

Using the new MCMC samples, we consider prediction of the Poisson rate (per
unit time) at 1709 locations covering the island using the four
candidate models and also the ensemble prediction given in~\eqref{eq:11}. These predictions are
shown in Figure~\ref{fig:rongelap_pred}, along with the observed count per
unit time. It can be seen that the
predicted Poisson rate has similar pattern for all models, with higher values at the
west side of the island, and matches that of the observed data closely. Examination of the range of prediction across
all locations shows that the exponential model has the highest range (4.9 to
10), followed by the exponential-power (5 to 9.9), the Mat{\'e}rn (5.1 to 9.9),
and the spherical model (5.4 to 9.8), while the ensemble model's range is
5.2 to 9.9. The prediction standard deviation falls in the range of 2.3 to
2.7 for all models.

To assess the predictive performance of each model, we performed
leave-one-out crossvalidation. For each $i=1,\ldots,n$, the observation
$y_i$ was deleted from the data set, and each model was fitted to the
remaining data using the same two-stage procedure described in the
beginning of this section. Let $\mu_i$ denote the mean per unit time at the
location of the deleted observation, $\ybf_{\setminus i}$ the vector of
observations without $y_i$, and $\hat{\xi}_{\setminus i}$ the estimate of
$\xi$ from the first stage using data $\ybf_{\setminus i}$. At the end of
the first stage, a MCMC sample $\mu_i^{(1)}, \ldots, \mu_i^{(L)}$, with
$L=1000$ after a burn in of 300 samples, was obtained from the conditional
distribution of $\mu_i$ given $(\ybf_{\setminus i}, \hat{\xi}_{\setminus i})$. We
evaluate each model, as well as the ensemble model, by calculating its
average negative predictive score given by
\begin{equation*}
  \mathrm{NegScore} = -\frac{1}{L} \sum_{l=1}^L \sum_{i=1}^n \log p[y_i |
  \mu_i^{(l)}], 
\end{equation*}
where $p[y_i | \mu_i]$ denotes the Poisson pmf with rate
$t_i \mu_i$ evaluated at $y_i$. The model with the lowest negative score
is preferred. We also calculate the average root mean square error given by
\begin{equation*}
  \mathrm{RMSE} = \sqrt{\frac{1}{L} \sum_{l=1}^L \sum_{i=1}^n (y_i - t_i
    \mu_i^{(l)})^2 }.
\end{equation*}
The model with the lowest RMSE is preferred. The results are shown in
Table~\ref{tab:rongelap_crossval}. The results show that the exponential
model has the best predictive performance, and the exponential-power model has
the worst performance. The weights given in Table~\ref{tab:ronge_est} agree with this
ranking. The ensemble model is significantly better than any of the
individual models.

\begin{table}
  \centering
  \begin{tabular}{lrrrrr}
    \hline
              & Mat\'ern & Exp-power & Spherical & Exponential & Ensemble \\ \hline
    NegScore  & 63976 & 64923 & 63896 & 62482 & 36797 \\ 
    RMSE      & 21359 & 21499 & 21406 & 21185 & 17329 \\
    \hline
  \end{tabular}
  \caption{Negative score and RMSE for the models used in the analysis of
    the Rongelap example. Smaller values are preferred.}
  \label{tab:rongelap_crossval}
\end{table}

\subsubsection{Comparison with a fully-Bayesian approach}
\label{sec:comp-FB}

An alternative to our EB method is a fully-Bayesian (FB) analysis. In
FB method, the components of $\xi$ are also assigned priors together
with the priors on $\psi$. Also, in this case, MCMC algorithms are
used to obtain samples from the joint posterior distribution of
$\zbf$, $\psi$ and $\xi$. Sampling from this posterior distribution
can be difficult because of the correlation between the parameters
\citep{chri:robe:skol:2006}. In this section we apply a FB approach to
the Rongelap data and compare it with the method proposed in this
paper. Since we have identified that the model with the exponential
correlation provides the best fit for these data, we focus on this
model. We also fix the link function parameter to its estimate
$\nu=0.957$, as the choice of an appropriate prior on this degrees of
freedom parameter is known to be problematic
(\citet[][p. 20]{doss:2012}, \citet[][p. 99-100]{roy:2014}).

Previously, we used Laplace approximation to identify a suitable range
for $\phi$ within $(178,975)$ and for $\omega$ within $(1.00,3.82)$
(see Figure~\ref{fig:rongelap_LA}). We consider two different FB
models depending on the choice of prior. Model FB1
assumes independent uniform priors using the information
of these ranges, i.e., $\pi(\phi) \propto 1_{(178,975)}(\phi)$ and
$\pi(\omega) \propto 1_{(1.00,3.82)}(\omega)$.  Mimicking the scale
invariant prior $\pi(\phi) \propto 1/\phi$, model FB2 assumes
$\pi(\phi) \propto \phi^{-1} 1_{(0,2000)}(\phi)$ and
$\pi(\omega) \propto \omega^{-1} 1_{(0,5)}(\omega)$.
The other parts of the model remained the same.

Using trial MCMC runs, we selected Metropolis-Hastings steps
(with joint updates for $(\phi,\omega)$) so that the
acceptance rate is about 25\%. A total of 55000 MCMC samples, after
a burn in of 300 samples, were selected from the posterior
distribution of $(\zbf,\psi,\xi)$ given the data. The total MCMC
sample size matches the one from the EB analysis.

\begin{figure}
  \centering
  \includegraphics[width=.9\linewidth]{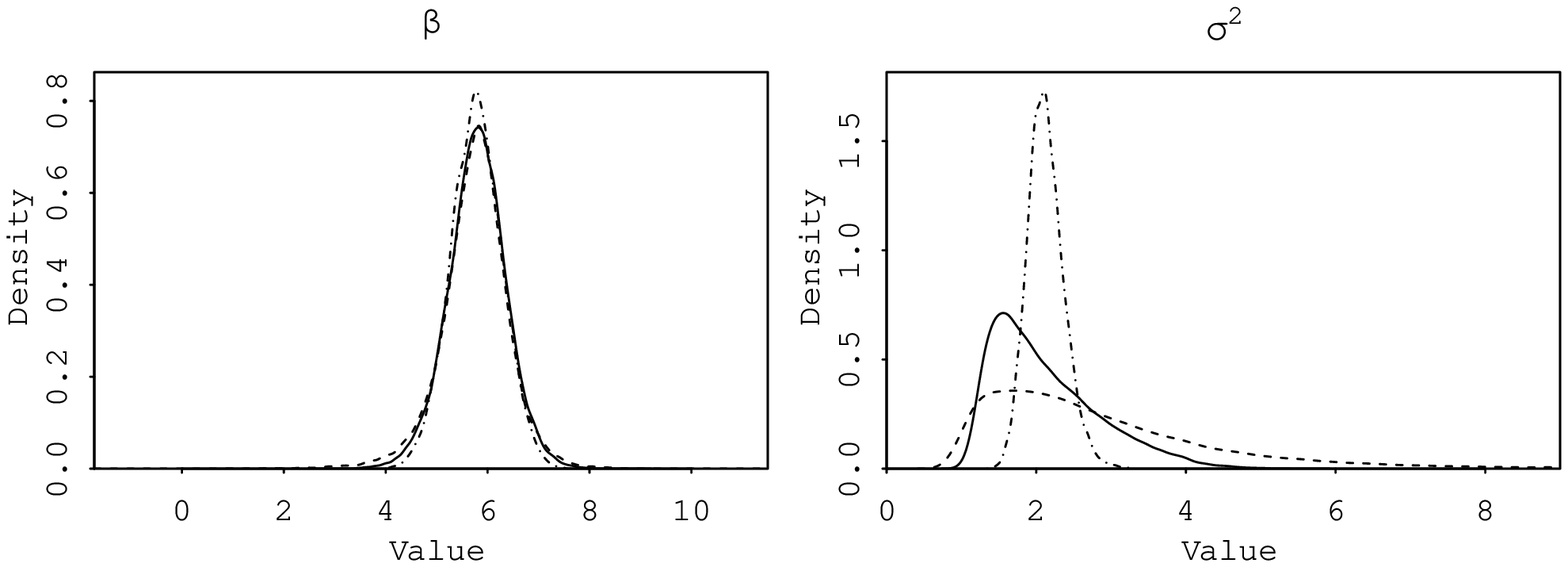} \\
  \includegraphics[width=.9\linewidth]{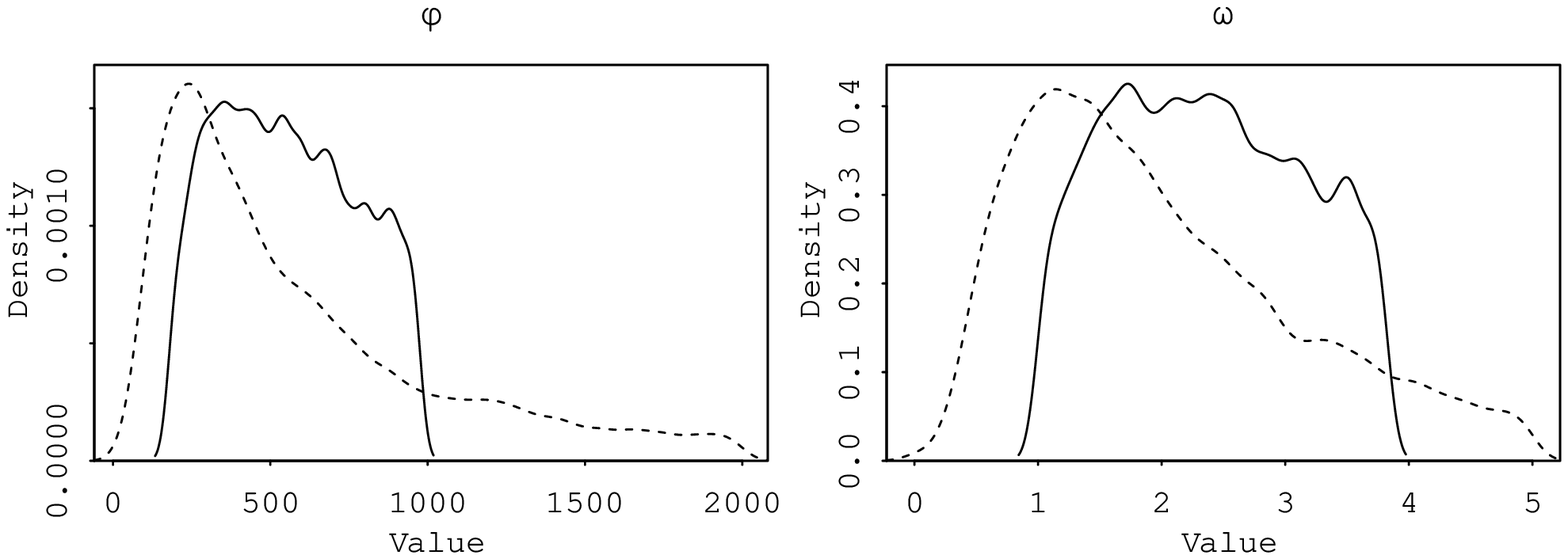}
  \caption{Posterior densities for the two fully-Bayesian approaches, FB1
    (solid line) and FB2 (dashed line),
    and the proposed empirical Bayes approach (dashed-dotted line) for the
    Rongelap example.}
  \label{fig:rongelap_FB1}
\end{figure}

\begin{figure}
  \centering
  \includegraphics[width=\linewidth]{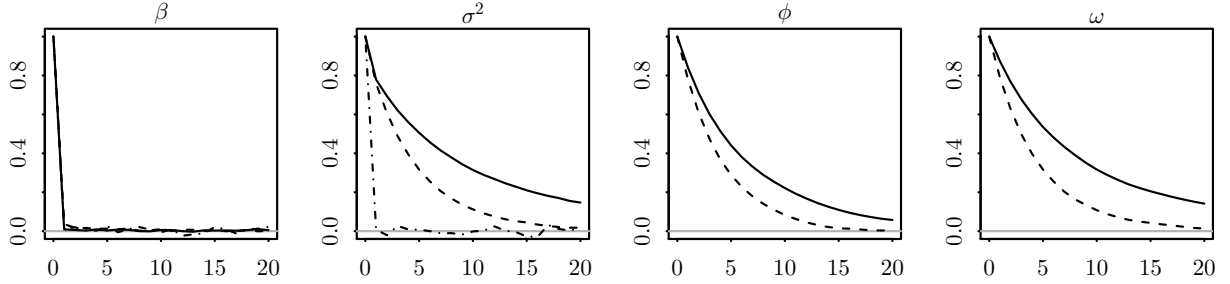}
  \caption{Autocorrelation plots of the MCMC samples for the two fully-Bayesian approaches, FB1
    (solid line) and FB2 (dashed line),
    and the proposed empirical Bayes approach (dashed-dotted line) for the
    Rongelap example.}
  \label{fig:rongelap_FB_acf}
\end{figure}

In terms of computing time, the FB methods were slower: FB1
took 89 seconds, FB2 took 106 seconds, and the proposed EB took 48
seconds but with the additional overhead of having to estimate
$\nu$. Plots of the posterior densities for the parameters using
each method are shown in Figure~\ref{fig:rongelap_FB1}.  Although
the posterior density for $\beta$ is similar using either method,
the posterior for $\sigma^2$ has higher variance with the two FB
approaches compared to the EB approach, which is not surprising
given that FB analysis also samples $\phi$ and
$\omega$. Furthermore, the posteriors for $\phi$ and $\omega$ in the
case of FB1 are not very informative, and are different from the
posteriors under FB2. This shows that the results are sensitive to
the choice of prior, something which the EB approach
avoids. The autocorrelation plots
(Figure~\ref{fig:rongelap_FB_acf}) show that the MCMC chains (except
for the $\beta$ chain) for the FB models suffer from high lag
covariances. One of the reasons for slow mixing of the Markov
chains in the FB models is the strong (posterior) correlation
between the parameters, for example, the correlation
between $\sigma^2$ and $\omega$ is $-0.9$ for FB1 and
$-0.8$ for FB2. The mixing of the MCMC algorithms for FB models may be improved by reparameterization \citep{chri:robe:skol:2006}, although it is unclear how this can be implemented for the general models
considered in this paper.

\subsubsection{Comparison with the untransformed estimator and separability
  of the Box-Cox model}
\label{sec:comp-with-unpar}

As discussed in Section~\ref{sec:gis}, if the skeleton set is chosen
sparsely, then the Monte-Carlo sample can become separable. This phenomenon
is particularly acute when the link function changes significantly for
small changes of its parameter, which is the case for the Box-Cox link. In
fact, for this model the MCMC sample can be \textit{completely} separable,
therefore the RL estimator~\eqref{eq:14} is unidentifiable when using the
untransformed samples. Consequently, the estimator~\eqref{eq:dhatbfsg} is
undefined and so is the EB estimate for $\xi$.

We focus on estimation of the link function parameter only by
maximizing the estimated BF's because the separability issue arises
when the link function parameter varies. We use the Poisson modified
Box-Cox model with exponential correlation, and fix the covariance
parameters at $\phi=400$ and $\omega = 2.2$. The prior distributions
for $\beta$ and $\sigma^2$ remain unchanged from our original
analysis of these data. The skeleton set for $\nu$ is set to
$\Xi = \{\xi_1, \xi_2, \xi_3\} = \{0.8, 1.0, 1.2\}$.  Despite these
values chosen to be very close, we will see that the untransformed
estimator fails to estimate the BF between these three models
accurately.

For each value of $\nu \in \Xi$, we take an MCMC sample from the posterior
distribution of $(\psi,\zbf)$ of length 1300 out of which the first
300 samples are discarded and the final $N=1000$ samples are retained. Let 
$(\psi^{(j;l)}, \zbf^{(j;l)})$ denote the $l$th sample from
$\pi_{\xi_j} (\psi, \zbf |\ybf)$ when $\nu = \xi_j$, for
$l=1,\ldots,N$, $j = 1,2,3$. Also
let $\mu^{(j;l)} = h_{\nu_j}^{-1}(\zbf^{(j;l)}), l=1,\ldots,N$ be
the transformed posterior samples for the mean, that is, from
$\pi_{\xi_j} (\psi, \mubf |\ybf)$ given in~\eqref{eq:postsimu}. The RL estimator can be
evaluated using either $\zbf^{(j;l)}$ or $\mu^{(j;l)}$ samples. The
quasi log likelihood, which is maximized to obtain the RL estimator is
defined in terms of the sample inclusion probabilities $\tilde{p}_j$'s
in~\eqref{eq:ptil} which in turn is defined in terms of the likelihood
\begin{equation*}
  L_\zbf^{(i;j;l)} = p[\ybf|\mubf = f_{\nu_i}(\zbf^{(j;l)})]
  p[\zbf^{(j;l)} | \psi^{(j;l)}, \nu_i],
\end{equation*}
when using $\zbf^{(j;l)}$ samples
and
\begin{equation*}
  L_\mu^{(i;j;l)} =
  p[h_{\nu_j}(\mu^{(j;l)}) | \psi^{(j;l)}, \nu_i] \tilde J_{\nu_i}(\mu^{(j;l)}),
\end{equation*}
when using $\mu^{(j;l)}$ samples. The separability issue discussed in
\pcite{geye:1994} arises if there exists a partition $\Xi_1,\ldots,\Xi_m$
of skeleton points $\Xi$ such as for each $(j;l)$, there exists
$r \in \{1,\ldots,m\}$ such that $\xi_i \notin \Xi_r$ implies
$L^{(i;j;l)} = 0$. In this case the Bayes factors can be estimated for
densities within the same partition but not between partitions. For the
chosen model, separability is mathematically impossible, but can happen
numerically if the corresponding observed (Markov chain) sample and the
Poisson rate parameter are very different.

\begin{figure}
  \centering
  \includegraphics[width=.8\linewidth]{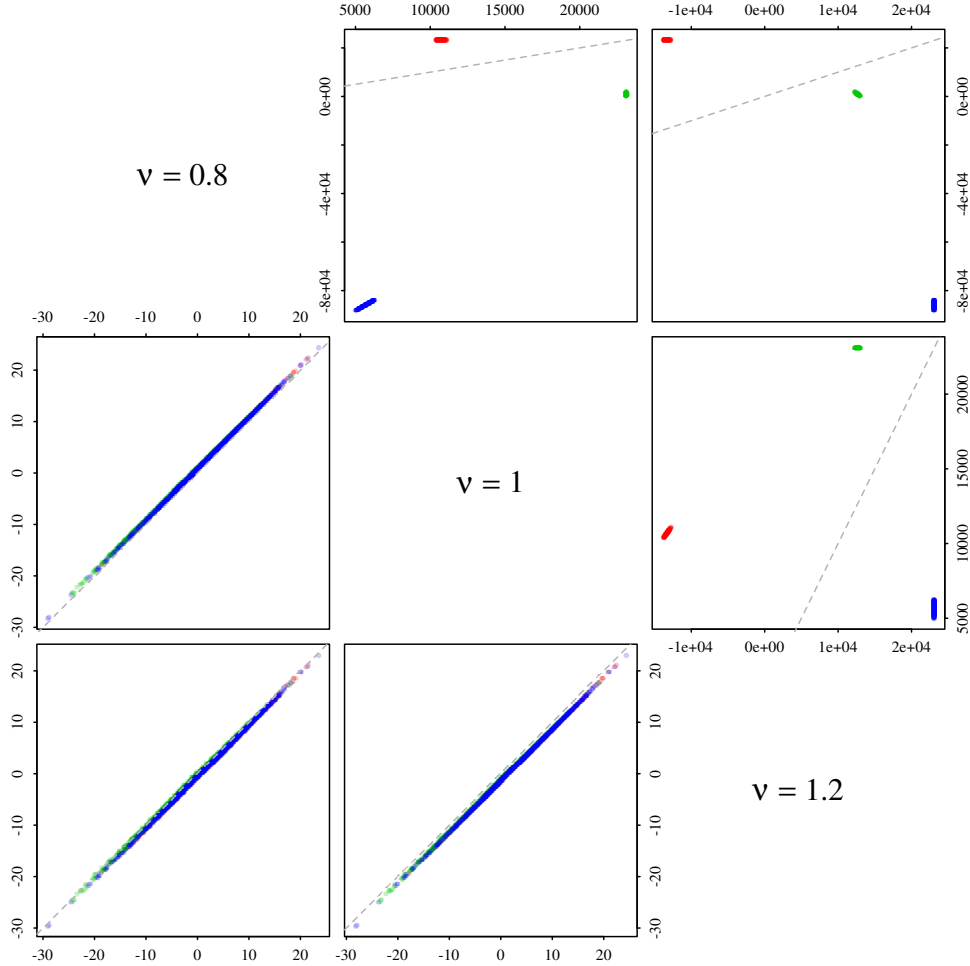}
  \caption{Log-likelihood values evaluated at different link parameters as
    indicated in the diagonal  evaluated at data generated from the
    posterior distribution with link parameter: $\nu = \xi_1$ (red),
    $\nu = \xi_2$ (green), $\nu = \xi_3$ (blue) and plotted against each other.
    The upper triangle shows the log-likelihoods for the untransformed
    samples and the lower triangle shows the log-likelihoods for the
    transformed samples. The dashed line corresponds to the line with slope 1 and intercept 0.}
  \label{fig:rongelap_sep}
\end{figure}

Figure~\ref{fig:rongelap_sep} shows plots of log-likelihood values
($\log L_\zbf^{(i;j;l)}$) plotted against $\log L_\zbf^{(i';j;l)}$ for
$i \neq i'$ in the upper triangle. The colors correspond to each $j$.
Similarly in the lower triangle we plot $\log L_\mu^{(i;j;l)}$ against
$\log L_\mu^{(i';j;l)}$. It can be seen (Table~\ref{tab:rongelap_sep})
that the $\log L_\zbf^{(i;j;l)}$ are very different for different $i$
and their differences are in the range of tens of thousands so
when taking exponentials, it will yield a zero. (The inclusion
probabilities $\tilde{p}_j$'s in~\eqref{eq:ptil} depend on the ratio
of likelihoods $L_\zbf^{(i;j;l)}$'s which is equivalent to exponential
of differences of log-likelihoods.) On the other hand, the differences
for the transformed sample are in the range of ones so the
reparameterized sample does not suffer from the separability issue. Thus, when
using the likelihood from the untransformed samples, $L_\zbf^{(i;j;l)}$, it
is impossible to estimate $\nu$ because the BF estimators are unidentifiable.

\begin{table}
  \centering
  \begin{tabular}{lccc}
    \hline
    &       1      &       2         &         3          \\
    \hline
    $1 - 2$&$( 1.2\mathrm{e}{+4} ,1.3\mathrm{e}{+4})$ & $(-2.3\mathrm{e}{+4}, -2.1\mathrm{e}{+4})$ & $( -9.3\mathrm{e}{+4}, -9.0\mathrm{e}{+4})$ \\
    $1 - 3$&$( 3.6\mathrm{e}{+4} ,3.7\mathrm{e}{+4})$ & $(-1.3\mathrm{e}{+4}, -1.0\mathrm{e}{+4})$ & $( -1.1\mathrm{e}{+5}, -1.1\mathrm{e}{+5})$ \\
    $2 - 3$&$( 2.4\mathrm{e}{+4} ,2.4\mathrm{e}{+4})$ & $( 1.0\mathrm{e}{+4},  1.1\mathrm{e}{+4})$ & $( -1.8\mathrm{e}{+4}, -1.7\mathrm{e}{+4})$ \\
    \hline \\[5pt]
    \hline
    &       1      &       2         &         3          \\
    \hline
    $1 - 2$&$(-1.0, -0.6)$ & $(-1.3, -0.6)$ & $(-1.0, -0.6)$ \\
    $1 - 3$&$( 0.1,  1.0)$ & $(-0.4,  0.9)$ & $( 0.1,  1.0)$ \\
    $2 - 3$&$( 1.1,  1.6)$ & $( 0.9,  1.5)$ & $( 1.2,  1.6)$ \\
    \hline
  \end{tabular}
  \caption{Top: Range of differences
    $\log L_\zbf^{(i;j;l)} - \log L_\zbf^{(i';j;l)}$ for $(i,i')$ shown in
    the rows for each $j$ shown in the columns. Bottom: The same for
    $\log L_\mu^{(i;j;l)} - \log L_\mu^{(i';j;l)}$.}
  \label{tab:rongelap_sep}
\end{table}

\subsection{Analysis of the incidence rates of the \textit{Rhizoctonia} root
  rot}
\label{sec:rhizoctonia}

In this example we analyze the root infection rates caused by
\textit{Rhizoctonia} fungi on wheat and barley. Data were collected at 100
locations where 15 plants were pulled out at each location and the total
number of crown roots and infected crown roots were counted. These data
were originally analyzed by \cite{zhan:2002} who used a binomial SGLMM with
logit link and spherical correlation. 
In this paper we consider four different models. The link function is
chosen among a robit or modified GEV link and the correlation function
is chosen among a spherical or exponential model. Thus
$\xi = \{\nu,\phi\}$. It is known that the robit link with about 7
degrees of freedom provides an excellent approximation to the logit
link. Thus \pcite{zhan:2002} model is (approximately) part of our models to choose from.

The spatial random field is assumed to have constant mean $\beta$ and
partial sill parameter $\sigma^2$ which are
assigned the normal and scaled-inverse-chi-square priors of
Section~\ref{sec:eb} with hyperparameter values $m_b=0$, $V_b=10$, $n_\sigma=4$, and $a_\sigma=10$. We also fix $\omega = 0$ as
we found that estimating this parameter along with the other parameters
results in serious overfit to the data.

For each model we choose the skeleton set by the method described in
Section~\ref{sec:skel} with $T=4$ and discard points that fall below 60\%
of the maximum marginal value. This procedure resulted in 8, 8, 9, and 9 skeleton
points for the models in Table~\ref{tab:rhiz_est} respectively (see
Appendix~\ref{sec:list-skeleton-points}, Table~\ref{tab:skel_rhizoctonia}).

For each model, we generate Markov Chain samples of size given by $N/k$
(rounded down), where $N=50000$ and $k$ is the number of skeleton points
for each model given in the previous paragraph, from
$\pi_\xi(\beta, \sigma^2, \zbf |\ybf)$, after discarding a burn in of 300
samples, corresponding to each point $\xi$ in the skeleton set. We use
approximately 80\% of these samples for the reverse logistic regression
estimation and the remaining 20\% of the samples to form GIS estimators and
estimate $\hat{\xi}$. These estimates are shown in Table~\ref{tab:rhiz_est}
along with the posterior mean estimates of $\beta$ and $\sigma^2$ based on
the density $\pi_{\hat{\xi}}(\beta, \sigma^2, \zbf |\ybf)$ using 5000 MC
samples from this density, after a burn-in of 300 samples. It can be
seen that the chosen models are simplified versions of the more general
fitted models. In the case of the robit link, the probit link is
selected, and in the case of the modified GEV link with exponential
correlation, the Gumbel (log-log) link is selected.

\begin{figure}
  \centering
  \includegraphics[width=\linewidth]{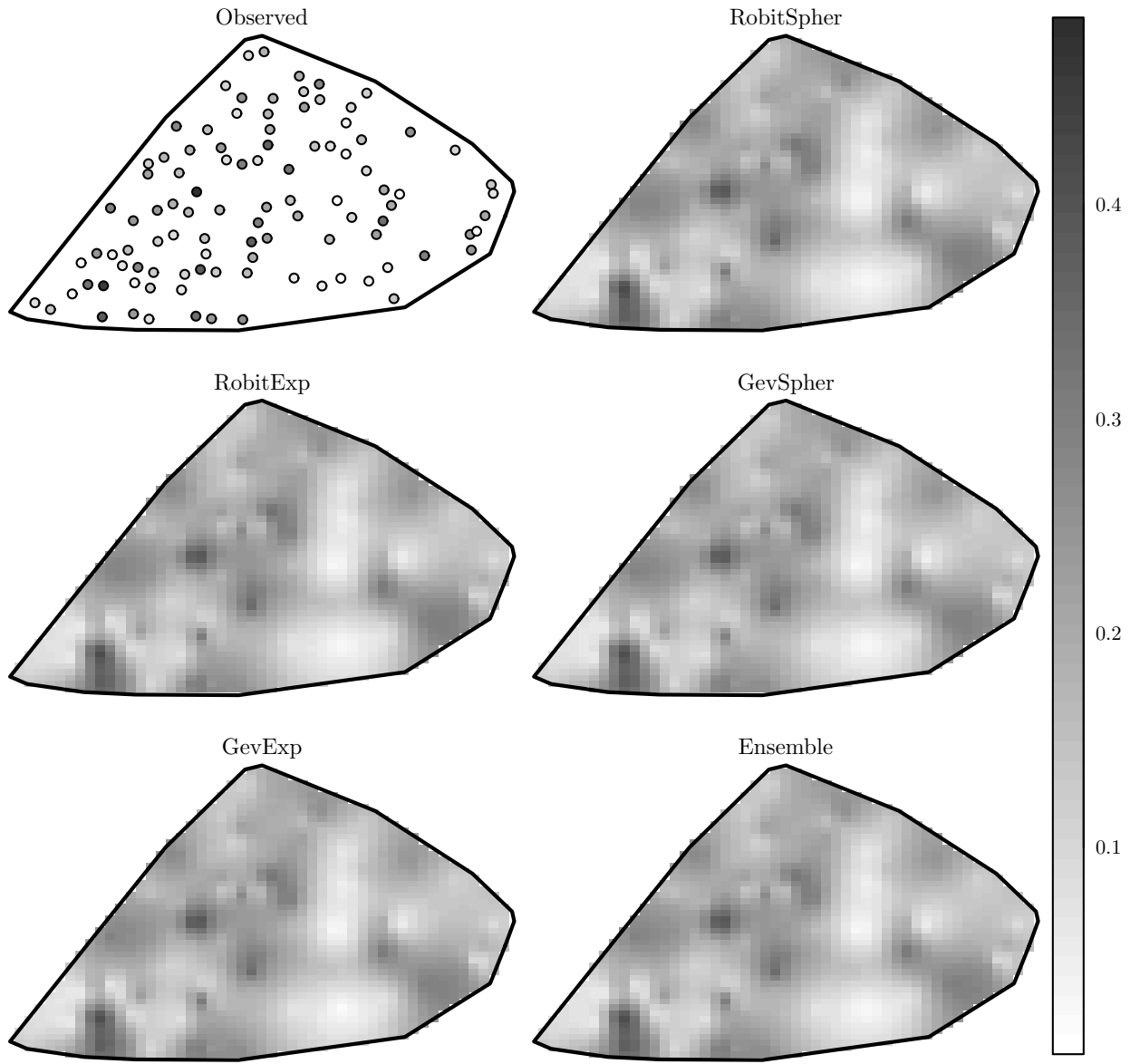}
  \caption{Observed proportion of infected roots and prediction of the binomial probability for the Rhizoctonia example under
    four different models and ensemble prediction.}
  \label{fig:rhizoctonia_pred}
\end{figure}

We calculate the weight of each model as discussed in
Section~\ref{sec:msel} using the samples generated from
$\pi_{\hat{\xi}}(\beta, \sigma^2, \zbf |\ybf)$. The estimates of the Bayes
factors relative to the probit-spherical model are shown in
Table~\ref{tab:rhiz_est}. It can be seen that the modified GEV link and
exponential correlation have higher weight than the robit link and
spherical correlation respectively. For most models the estimate of the
link function parameter is at the boundary of the parameter space so there
is evidence that using a parameterized link function is overfitting the
data.
The estimated infection probability (posterior mean) is shown in
Figure~\ref{fig:rhizoctonia_pred} from each model, along with the ensemble
prediction and the observed proportion of infections. It can be seen that the models give similar predictions and the
prediction pattern resembles that of \citet{zhan:2002} and of the observed data.

\begin{table}
  \centering 
  \begin{tabular}{llllllll}
    \hline
Model & $\beta$& $\sigma^2$& $\nu$ & $\phi$ & log BF
    & $|\xi|$ & Weight \\
    \hline
Robit, Spherical            & $-$0.983 & 8.232   & $\infty$ & 3113   & 0     & 1 & 0.192 \\
                            & (2.344)  & (1.448) &          & (2063) &       &   &       \\
Robit, Exponential          & $-$0.997 & 7.299   & $\infty$ & 1848   & 0.246 & 1 & 0.245 \\
                            & (2.277)  & (1.201) &          & (1248) &       &   &       \\
Modified-GEV, Spherical     & $-$0.619 & 8.275   & 0.067    & 4367   & 0.660 & 2 & 0.136 \\
                            & (2.525)  & (1.388) & (0.327)  & (3332) &       &   &       \\
Modified-GEV, Exponential   & $-$0.620 & 7.710   & 0        & 2922   & 0.801 & 1 & 0.427 \\
                            & (2.428)  & (1.312) &          & (1946) &       &   &       \\
    \hline
  \end{tabular}
  \caption{Parameter estimates and log Bayes factor relative to the
    Mat\'ern model for the Rhizoctonia example for each model. The size of
    $\xi$ is denoted by $|\xi|$ (counting only the components whose
    estimates fall in the
    interior of the parameter space).}
  \label{tab:rhiz_est}
\end{table}

A leave-one-out
crossvalidation was performed to assess each model fitted as well as the
ensemble model. For each model we calculate the average negative predictive
score as with the example of Section~\ref{sec:rongelap} but using the
binomial instead of the Poisson pmf, and the average RMSE. The results are
shown in Table~\ref{tab:rhizoctonia_crossval}. The results are conflicting:
the robit-spherical model is the best according to the negative score
criterion and the modified-GEV-exponential the worse, but the
modified-GEV-spherical is the best according to the RMSE criterion and the
robit-exponential the worse. There is a significant amount of variability
which makes it difficult to make a proper assessment among the four models
considered, however, it can be seen that the ensemble model, as with the
previous example, is again significantly better using either measure.

\begin{table}
  \centering
  \begin{tabular}{lrrrrr}
    \hline
    & Robit, Spher & Robit, Exp &  M-GEV, Spher
    &    M-GEV, Exp &  Ensemble \\ \hline
    NegScore & 1367 & 1371 & 1373 & 1381 & 956 \\
    RMSE     &   207 &   207 &   201 &   202 &  167 \\ \hline
  \end{tabular}
  \caption{Negative score and RMSE for the models used in the analysis of
    the rhizoctonia example. Smaller values are preferred.}
  \label{tab:rhizoctonia_crossval}
\end{table}

\section{Conclusion and discussion}
\label{sec:disc}

In this paper we discuss SGLMMs where the link function contains
unknown parameters. These models can be more robust compared to models
which use a prescribed link function. Some of the proposed flexible
link functions in the literature are not consistent with the Gaussian
assumption of the latent spatial field, so we propose simple
modifications to make them consistent without losing their
flexibility. The central theme of the paper is the estimation of the
link function and spatial correlation parameters by maximizing the
Bayes factors relative to a fixed model.  Therefore the methodology is
developed around the ability to compute these Bayes factors
efficiently. We show that naive generalized importance sampling
estimation can sometimes fail, and show how by using suitable
transformations to the samples can give accurate results. Thus we
develop effective GIS and reverse logistic estimators based on
appropriately chosen reparameterizations. The reparameterization is
shown to reduce the variability in GIS estimators, and also alleviates
the well-known separability problem of \pcite{geye:1994} reverse
logistic regression estimator. We also use the RL method to compare
models which have different families of link and correlation
functions, thus providing a method of choosing and weighting different
spatial models. This also allows for ensemble estimation and
prediction of the mean response. In fact, for the two examples presented in
this paper, the ensemble prediction outperforms predictions based on a
single model.

The choice of importance sampling densities can impact the accuracy of
the GIS estimators. In the context of the simple IS estimator,
\cite{bote:lecu:tuff:2013} discuss construction of semi-parametric and
nonparametric importance sampling densities using Markov chain samples
\cite[see also][]{beau:cald:2013}. Here, we use Laplace approximation
to marginal likelihoods for choosing suitable importance distributions
for the GIS estimators. The new reparameterized GIS estimators and the
EB methodology for selecting models, although developed in the context
of SGLMMs here, are applicable to other models including generalized
linear models and generalized linear mixed models. Also, the use of
transformation can be similarly extended to improve other IS
estimators, e.g.  other multiple IS schemes \citep{veac:guib:1995,
  owen:zhou:2000, elvi:mart:luen:buga:2015, mart:elvi:luen:cora:2017},
parallel, serial or simulated tempering \citep{geor:doss:2017,
  mari:pari:1992}. Likewise, the proposed method of choosing
importance densities for GIS can also be used for other IS estimators.

\appendix

\topsection{Appendices}

\section{Detailed derivations}
\label{sec:appendixA}

\subsection{Standard errors for empirical Bayes estimates}
\label{sec:apdx:sterr}

To estimate the variability in the empirical Bayes estimates for $\xi$, we
compute
\begin{equation*}
  \frac{\partial^2}{\partial\xi \partial\xi^\T} \log m_\xi(\ybf) = \E
  \left( \frac{\partial^2}{\partial\xi \partial\xi^\T} \log
    p[\ybf,\wbf|\psi,\xi] \right) + \Var \left(
    \frac{\partial}{\partial\xi} \log p[\ybf,\wbf|\psi,\xi] \right) ,
\end{equation*}
\citep[see][]{case:2001} where the expectation and variance are
taken with respect to the posterior density $\pi_\xi(\wbf,\psi|\ybf)$.

To derive an explicit formula for our model, we write
\begin{equation}
  \label{eq:8}
  \log p[\ybf|\mubf = f_\nu(g_\nu(\wbf))] = \sum_{i=1}^n \Bigg[\frac{1}{\chi}
  (y_i \gamma_i - t_i K(\gamma_i)) + c(y_i,\chi)\Bigg],
\end{equation}
where $\gamma$ denotes the canonical parameter, $\chi$ is the dispersion
parameter which is assumed known, $K(\gamma)$ is the cumulant function such
that $K'(\gamma) = \mu$, and $c(y,\chi)$ is a function which does not
depend on $\gamma$ and not relevant to our analysis. For the
binomial and Poisson models discussed here, $K(\gamma) = \log(1+e^\gamma)$
and $K(\gamma) = \exp(\gamma)$ respectively, and in both cases $\chi=1$
\citep{McCu:Neld:gene:1999}. Specifically, we have the following
relationship between $\gamma_i$ and $w_i$,
$K'(\gamma_i) = f_\nu(g_\nu(w_i))$.

We also write $\vartheta = \{\theta,\omega\}$ for the covariance parameters and
\begin{equation}
  \label{eq:9}
  \log p[\zbf = g_\nu(\wbf)|\psi,\xi] = -\frac{1}{2\sigma^2} (\zbf - X\beta)^\T
  R_{\vartheta}^{-1} (\zbf - X\beta)
  - \half \log |R_{\vartheta}| - \frac{n}{2} \log(2\pi\sigma^2) ,
\end{equation}
where $R_{\vartheta}$ denotes the matrix whose $(i,j)$ element is
$\rho_\theta(\|s_i-s_j\|) + \omega I_{\{s_i = s_j\}}$ for sampling locations
$s_i$, $s_j$, $i,j=1,\ldots,n$.

We now proceed to compute the necessary derivatives from~\eqref{eq:8}
and~\eqref{eq:9}. Note that
\begin{align*}
  \frac{\partial}{\partial\nu} \log p[\ybf|\mubf]
  &= \frac{1}{\chi} \sum_{i=1}^n (y_i - t_i K'(\gamma_i)) \frac{\partial
    \gamma_i}{\partial \nu} , \\
  \frac{\partial^2}{\partial\nu^2} \log p[\ybf|\mubf]
  &= \frac{1}{\chi} \sum_{i=1}^n (y_i - t_i K'(\gamma_i)) \frac{\partial^2
    \gamma_i}{\partial \nu^2} - \frac{1}{\chi} \sum_{i=1}^n t_i
    K''(\gamma_i) \left(\frac{\partial \gamma_i}{\partial \nu} \right)^2 ,
\end{align*}
where
\begin{align*}
  K''(\gamma_i) \frac{\partial \gamma_i}{\partial \nu}
  &= \frac{\partial}{\partial \nu} f_\nu(z_i) \cdot
    \frac{\partial}{\partial \nu} g_\nu(w_i), \\
  K''(\gamma_i) \frac{\partial^2 \gamma_i}{\partial \nu^2} + K'''(\gamma_i)
  \left( \frac{\partial \gamma_i}{\partial \nu} \right)^2
  &= \frac{\partial^2}{\partial \nu^2} f_\nu(z_i) \cdot
    \left(\frac{\partial}{\partial \nu} g_\nu(w_i) \right)^2 +
    \frac{\partial}{\partial \nu} f_\nu(z_i) \cdot
    \frac{\partial^2}{\partial \nu^2} g_\nu(w_i) .
\end{align*}
From~\eqref{eq:9} we have
\begin{align*}
  \frac{\partial}{\partial \nu} \log p[\zbf|\psi,\xi]
  &= -\frac{1}{\sigma^2} (\zbf - X\beta)^\T R_{\vartheta}^{-1}
    \left( \frac{\partial}{\partial \nu} g_\nu(\wbf) \right) , \\
  \frac{\partial^2}{\partial \nu^2} \log p[\zbf|\psi,\xi]
  &= -\frac{1}{\sigma^2} \left( \frac{\partial}{\partial \nu} g_\nu(\wbf)
    \right)^\T R_{\vartheta}^{-1}
    \frac{\partial}{\partial \nu} g_\nu(\wbf) -\frac{1}{\sigma^2} (\zbf -
    X\beta)^\T R_{\vartheta}^{-1}
    \frac{\partial^2}{\partial \nu^2} g_\nu(\wbf) .
\end{align*}

We write $\partial_j R_{\vartheta}$ for the derivative of
$R_\vartheta$ with respect to the $j$th component of $\vartheta$ and
similarly for higher-order derivatives. We have,
\begin{align*}
  \frac{\partial}{\partial \vartheta_j} \log p[\zbf|\psi,\xi]
  ={}& \frac{1}{2\sigma^2} (\zbf - X\beta)^\T (R_\vartheta^{-1} \partial_j R_{\vartheta}
    R_\vartheta^{-1}) (\zbf - X\beta) - \half \mathrm{tr} (R_\vartheta^{-1} \partial_j
    R_{\vartheta}) ,\\
  \frac{\partial^2}{\partial \vartheta_j \partial \vartheta_k} \log
  p[\zbf|\psi,\xi]
  ={}& \frac{1}{2\sigma^2} (\zbf - X\beta)^\T (R_\vartheta^{-1} \partial^2_{jk} R_{\vartheta}
    R_\vartheta^{-1}) (\zbf - X\beta) \\
    & {} - \frac{1}{\sigma^2} (\zbf - X\beta)^\T (R_\vartheta^{-1} \partial_j
      R_{\vartheta}
    R_\vartheta^{-1} \partial_k R_{\vartheta}
    R_\vartheta^{-1}) (\zbf - X\beta)
  \\
    & {} + \half \mathrm{tr} (R_\vartheta^{-1} \partial_{j}
      R_{\vartheta} R_\vartheta^{-1} \partial_{k} R_{\vartheta}) - \half
      \mathrm{tr} (R_\vartheta^{-1} \partial^2_{jk} R_{\vartheta}) .
\end{align*}

Recall also the Jacobian term $\bar J_\nu(\wbf) = \prod_{i=1}^n
g'_\nu(w_i)$. Therefore,
\begin{align*}
  \frac{\partial}{\partial \nu} \log \bar J_\nu(\wbf)
  &= \sum_{i=1}^n
    \frac{1}{g'_\nu(w_i)} \frac{\partial}{\partial \nu} g'_\nu(w_i) , \\
  \frac{\partial^2}{\partial \nu^2} \log \bar J_\nu(\wbf)
  &= \sum_{i=1}^n
    \frac{1}{g'_\nu(w_i)} \frac{\partial^2}{\partial \nu^2} g'_\nu(w_i) -
    \sum_{i=1}^n  \left( \frac{1}{g'_\nu(w_i)} \frac{\partial}{\partial
    \nu} g'_\nu(w_i) \right)^2 .
\end{align*}

In practice we let $\mathcal{H}$ be the matrix
$\mathcal{H} = -\frac{\partial^2}{\partial\xi \partial\xi^\T} \log
m_{\hat\xi}(\ybf)$ and $\hat{\mathcal{H}}$ be its Monte-Carlo approximation
derived using samples $\{\zbf^{(l)},\psi^{(l)}\}_{l=1}^N$ from the
posterior density $\pi_{\hat{\xi}}(\zbf,\psi|\ybf)$ (or equivalently using samples
$\{g_{\hat\nu}^{-1}(\zbf^{(l)}),\psi^{(l)}\}_{l=1}^N$ from the
posterior density $\pi_{\hat{\xi}}(\wbf,\psi|\ybf)$).
Then, we approximate the variance of $\hat{\xi}$ by
$\hat{\mathcal{H}}^{-1}$.

\subsection{Laplace approximation}
\label{sec:apdx:lapl-appr}

We write the prior pdf for $\beta|\sigma^2$, $\pi(\beta|\sigma^2)$ as
\begin{equation*}
  \log \pi(\beta|\sigma^2) = -\frac{1}{2\sigma^2} (\beta - m_b)^\T V_b^{-1}
  (\beta - m_b) - \half \log |V_b| - \frac{p}{2} \log(2\pi\sigma^2) .
\end{equation*}
Simple calculations show that integrating out $\beta$,
$p[\zbf|\sigma^2,\xi] = \int_{\mathcal{R}^p} p[\zbf|\beta,
\sigma^2,\xi] \pi(\beta|\sigma^2) \ud\beta$ is given by
\begin{equation}
  \label{eq:10}
  \log p[\zbf|\sigma^2,\xi] = -\frac{1}{2\sigma^2} (\zbf - X m_b)^\T
  T_{\vartheta} (\zbf - Xm_b)
  + \half \log |T_{\vartheta}| - \frac{n}{2} \log(2\pi\sigma^2) ,
\end{equation}
where
\begin{equation*}
  T_{\vartheta} = R_{\vartheta}^{-1} - R_{\vartheta}^{-1} X
  (V_b^{-1} + X^\T R_{\vartheta}^{-1} X)^{-1} X^\T R_{\vartheta}^{-1} .
\end{equation*}
Then, from~\eqref{eq:8} and~\eqref{eq:10}, we choose $\tilde{\zbf}$ such that
\begin{equation*}
  \tilde{\zbf} = \tilde{\zbf}_\xi(\sigma^2) =\argmax_\zbf \log
  p[\ybf|\zbf,\xi] + \log p[\zbf|\sigma^2,\xi] ,
\end{equation*}
which is straightforward to obtain using a quasi-Newton algorithm
\citep{byrd:1995}.

The matrix $\tilde{H}_\xi(\sigma^2)$ in \eqref{eq:htil} is given by
\begin{equation*}
  \tilde{H}_\xi(\sigma^2) = \frac{1}{\sigma^2} T_\vartheta + \frac{1}{\chi}
  \tilde{D}_\vartheta ,
\end{equation*}
where $\chi$ is as in Appendix~\ref{sec:apdx:sterr} and
\begin{equation*}
  \tilde{D}_\vartheta = \mathrm{diag}
  \left\{
    t_i f'_\nu(\tilde{z}_i) \left. \frac{\partial \gamma_i}{\partial
        z_i} \right|_{z_i = \tilde{z}_i} - (y_i - t_i
  f_\nu(\tilde{z_i})) \left. \frac{\partial^2 \gamma_i}{\partial
        z_i^2} \right|_{z_i = \tilde{z}_i}
  \right\}_{i=1}^n ,
\end{equation*}
which is used for the evaluation of the integrand in~\eqref{eq:LAmarg}.

\section{Summary of the steps involved in inference}
\label{sec:summ}

In the proposed empirical Bayes formulation for SGLMMs, first, an
estimate of $\hat\xi$ is found using one of the reparameterized GIS
methods that involves the following two stages.

\noindent\rule{\linewidth}{0.4pt}
\begin{description}
\item[Stage 1] Draw MCMC samples
  $\{\psi^{(j;l)}, \zbf^{(j;l)}\}_{l=1}^{\tilde{N}_j}$ from
  $\pi_{\xi_j}(\psi,\zbf|\ybf)$ for $j=1,\ldots,k$, and use these
  samples to estimate ${\rbf}$ by the reverse logistic regression
  method. For the log quasi likelihood function in the RL estimation,
  $p[\ybf, \zbf|\psi, \xi_j]$ can be replaced by either
  $p[\zbf = h_{\nu_j} (\mubf) | \psi, \xi_j] \tilde J_{\nu_j}(\mubf)$
  or
  $p[\ybf|\mubf = f_{\nu_j}( g_{\nu_j} (\wbf))] p[\zbf = g_{\nu_j}
  (\wbf) | \psi, \xi_j] \bar J_{\nu_j}(\wbf)$ depending on whether the
  reparameterized samples are obtained using the transformation
  $h_\nu^{-1}: \zbf \mapsto \mubf$ or $g^{-1}_\nu: \zbf \mapsto \wbf$.

\item[Stage 2] Independently of Stage~1, draw new MCMC samples
  $\{\psi^{(j;l)}, \zbf^{(j;l)}\}_{l=1}^{N_j}$ from
  $\pi_{\xi_j}(\psi,\zbf|\ybf)$ for $j=1,\ldots,k$. Use these samples and
  $\hat {\bf r}$ computed in Stage 1 to estimate the BFs $B_{\xi, \xi_1}$
  by either of the two proposed reparameterized GIS estimators
  $\tilde{B}_{\xi, \xi_1} (\tilde{\rbf})$ (given in \eqref{eq:glsbf2}) or
  $\bar{B}_{\xi, \xi_1} (\bar{\rbf})$ (given in \eqref{eq:glsbf3}).

\item[] Estimate $\hat{\xi}$ by maximizing either
  $\tilde{B}_{\xi, \xi_1} (\tilde{\rbf})$ or
  $\bar{B}_{\xi, \xi_1} (\bar{\rbf})$.
\end{description}
\nointerlineskip
\noindent\rule{\linewidth}{0.4pt}
\vspace{\lineskip}

After finding the EB estimate $\hat{\xi}$, draw new MCMC samples
$\{\psi^{(i)}, \zbf^{(i)}\}_{i=1}^{M}$ from
$\pi_{\hat{\xi}}( \psi, \zbf| \ybf)$ to make inference on $\psi$ as
well as the latent Gaussian random field $\{Z(s), s \in \Sbb\}$. If
multiple families of link functions (and/or covariance functions) are
under consideration, then the ensemble estimates given in
\eqref{eq:11} can be used to make inference on $\psi$ and the random
field.

\section{List of skeleton points used in the examples}
\label{sec:list-skeleton-points}

This section lists the skeleton set obtained using the method of
Section~\ref{sec:skel} for the examples of Section~\ref{sec:rongelap}
(Table~\ref{tab:rongelap_skel}) and Section~\ref{sec:rhizoctonia}
(Table~\ref{tab:skel_rhizoctonia}). 

\begin{table}[h]
  \centering
  \begin{tabular}{rrrrrrrrrrrrrr}
    \toprule
    \multicolumn{4}{c}{Mat\'ern}
    & \multicolumn{4}{c}{Exp-power}
    & \multicolumn{3}{c}{Spherical}
    & \multicolumn{3}{c}{Exponential} \\
    \cmidrule(r){1-4} \cmidrule(lr){5-8} \cmidrule(lr){9-11}
    \cmidrule(l){12-14}
    $\nu$ & $\phi$ & $\omega$ & $\kappa$ & $\nu$ & $\phi$ & $\omega$ &
    $\kappa$ & $\nu$ & $\phi$ & $\omega$ & $\nu$ & $\phi$ & $\omega$ \\
  0.94 & 415 & 0.970 & 0.28 & 0.96 &  140 & 0.770 & 0.410 & 0.97 &  660 & 2.65 & 0.96 & 580 & 2.4 \\
  0.94 & 700 & 0.970 & 0.28 & 0.96 &  720 & 0.770 & 0.410 & 0.97 & 1130 & 2.65 & 1.10 & 580 & 2.4 \\
  0.94 & 415 & 2.385 & 0.28 & 0.96 & 1300 & 0.770 & 0.410 & 1.10 & 1130 & 2.65 & 0.96 & 980 & 2.4 \\
  0.94 & 700 & 2.385 & 0.28 & 0.96 &  720 & 2.035 & 1.005 & 0.97 & 1600 & 2.65 & 0.96 & 580 & 3.8 \\
  1.10 & 700 & 2.385 & 0.28 & 1.10 &  720 & 2.035 & 1.005 & 0.97 & 1130 & 4.30 &      &     &     \\
  0.94 & 130 & 2.385 & 0.94 & 0.96 &  720 & 3.300 & 1.005 &      &      &      &      &     &     \\
  0.94 & 415 & 2.385 & 0.94 & 1.10 &  720 & 3.300 & 1.005 &      &      &      &      &     &     \\
  1.10 & 415 & 2.385 & 0.94 & 0.96 &  720 & 2.035 & 1.600 &      &      &      &      &     &     \\
  0.94 & 415 & 3.800 & 0.94 & 0.96 &  720 & 3.300 & 1.600 &      &      &      &      &     &     \\
  1.10 & 415 & 3.800 & 0.94 & 1.10 &  720 & 3.300 & 1.600 &      &      &      &      &     &     \\
  0.94 & 130 & 2.385 & 1.60 &      &      &       &       &      &      &      &      &     &     \\
  0.94 & 130 & 3.800 & 1.60 &      &      &       &       &      &      &      &      &     &     \\
    \bottomrule
  \end{tabular}
  \caption{Skeleton set for Section~\ref{sec:rongelap}.}
  \label{tab:rongelap_skel}
\end{table}

\begin{table}[h]
  \centering
  \begin{tabular}[b]{cccc}
    \toprule
    Robit, spherical
    & Robit, exponential
    & Mod GEV, spherical
    & Mod GEV, exponential \\
    \cmidrule(r){1-1} \cmidrule(lr){2-2} \cmidrule(lr){3-3} \cmidrule(l){4-4}
    \begin{tabular}[t]{rr}
 $\nu$&$\phi$ \\ 
    12& 3300  \\ 
    21& 3300  \\ 
    31& 3300  \\ 
    40& 3300  \\ 
    21& 4900  \\ 
    31& 4900  \\ 
    40& 4900  \\ 
    40& 6500  
    \end{tabular}
    &
      \begin{tabular}[t]{rr}
$\nu$ &$\phi$ \\  
   40 & 990   \\  
   21 &1960   \\                         
   30 &1960   \\  
   40 &1960   \\  
   21 &2930   \\  
   30 &2930   \\  
   40 &2930   \\  
   40 &3900    
      \end{tabular}
    &
      \begin{tabular}[t]{rr}
$\nu$ &$\phi$  \\ 
 0.12 &2400    \\ 
 0.23 &2400    \\ 
 0.35 &2400    \\ 
 0.00 &4667    \\ 
 0.12 &4667    \\ 
 0.23 &4667    \\ 
 0.00 &6933    \\        
 0.12 &6933    \\ 
 0.00 &9200 
      \end{tabular}
    &
      \begin{tabular}[t]{rr}
 $\nu$ &$\phi$ \\ 
  0.12 &1400 \\   
  0.25 &1400 \\   
  0.37 &1400 \\   
  0.00 &2767 \\   
  0.12 &2767 \\   
  0.25 &2767 \\   
  0.00 &4133 \\   
  0.12 &4133 \\   
  0.00 &5500
      \end{tabular}
         \\
        \bottomrule
  \end{tabular}
  \caption{Skeleton set for Section~\ref{sec:rhizoctonia}.}
  \label{tab:skel_rhizoctonia}
\end{table}

\section*{Acknowledgment}
The authors thank two anonymous reviewers and an editor for several
helpful comments and suggestions that led to a substantially improved
revision of the paper.

\bibliographystyle{apalike}
\bibliography{ref}

\end{document}